\documentclass[11pt,a4paper]{article}

\pdfoutput=1

\usepackage{cite}
\usepackage{jheppub}
\usepackage{graphicx,bm,color}
\usepackage{slashed,verbatim}
\usepackage{caption}
\usepackage{subcaption}
\usepackage{amssymb,graphicx,epstopdf}
\usepackage[colorlinks=true,linktocpage=true,linkcolor=blue,citecolor=blue]{hyperref}
\def\be {\begin{equation}}
\def\ee {\end{equation}}
\def\bea {\begin{eqnarray}}
\def\eea {\end{eqnarray}}
\def\bc {\begin{center}}
\def\ec {\end{center}}
\def\nn {\nonumber}

\DeclareGraphicsExtensions{.jpg,.pdf,.eps}

\begin{document}
\title{Power corrections  to the electromagnetic spectral function and the dilepton rate 
in QCD plasma within operator product expansion in $D=4$}

\author{Aritra Bandyopadhyay and Munshi G Mustafa}
 \affiliation{
 Theory Division, Saha Institute of Nuclear Physics, HBNI, 
 1/AF Bidhan Nagar, Kolkata 700064, India.}
\emailAdd{aritra.bandyopadhyay@saha.ac.in  munshigolam.mustafa@saha.ac.in}

\abstract{
We evaluate the electromagnetic spectral function in QCD plasma in a nonperturbative
background of in-medium quark and gluon condensates by incorporating the leading order 
power corrections in a systematic framework within the ambit of the operator product 
expansion in $D=4$ dimension. We explicitly show that the mixing of the composite 
operators removes  mass singularities and renders Wilson coefficients finite and well 
defined. As a spectral property, we then obtain the nonperturbative dilepton  
production rate from QCD plasma. The operator product expansion automatically restricts 
the dilepton rate to the intermediate mass range,  which is 
found to be enhanced due to the power corrections. We also compare our result with 
those from  nonperturbative calculations, \textrm{e.g.}, lattice QCD and effective 
QCD models based on Polyakov loop.}

\maketitle

\section{Introduction}

Over the last couple of decades, with the international efforts from the relativistic 
heavy-ion 
collision experiments 
in SPS to LHC, we already have some profound signatures of the high temperature 
deconfined 
phase of quantum chromodynamics (QCD), namely quark gluon plasma (QGP). However, a 
locally 
equilibrated plasma is short-lived in the collision. However,  there 
are always some initial or final state interactions that may contaminate an observable   
one is interested in. In this respect the electromagnetic emissivity of the plasma in the 
form of real or virtual photon is 
particularly important. The very fact that they do not suffer from final state 
interactions and carries least contaminated information of the local equilibrium makes 
real or virtual photon production a desirable candidate for studying QGP. Real photon 
escapes unperturbed and virtual photon decays into a lepton pair in the process. This is 
why, the photon and dilepton production rates from QGP phase have been studied vividly in 
the last three decades~\cite{Lebellac, Weldon:1983jn, Weldon:1990iw, McLerran:1984ay, 
Hwa:1985xg, Kajantie:1986dh, Kajantie:1986cu, Cleymans:1986na,Cleymans:1992gb, 
Gale:1987ey, Gale:1987ki,Braaten:1990wp, Karsch:2000gi, Bandyopadhyay:2015wua, 
Greiner:2010zg, Aurenche:1998nw, Mustafa:1999dt, Karsch:2001uw, Ding:2010ga, 
Ding:2016hua, Islam:2014sea, Gale:2014dfa, Islam:2015koa, Bandyopadhyay:2016fyd, 
Sadooghi:2016jyf, Tuchin:2013bda, Srivastava:2002ic, Kvasnikova:2001zm,Chatterjee:2007xk, 
Laine:2013vma, Ghisoiu:2014mha, Ghiglieri:2014kma}. 

Even though the lepton pairs behave as free particles after production, but they are 
produced in every stage of the collisions. 
The high mass dileptons are mostly produced due to collision between 
hard partons and not particularly very informative about QGP. This is because the 
Drell-Yan processes~\cite{Drell:1969km} and charmonium decays~\cite{Dominguez:2009mk, 
Dominguez:2010mx} are the major 
processes in that regime. On the other hand the low mass dilepton production 
is enhanced~\cite{Adare:2009qk} compared to  all known sources of electromagnetic decay
of the hadronic particles and the contribution of a radiating
QGP.  So, the low mass dileptons ($\le 1$ GeV) possibly indicates some 
nonhadronic sources and the intricacies are discussed in the literature in a more 
phenomenological way~\cite{Greiner:2010zg,Islam:2014sea,Gale:2014dfa}. 
There also exists 
dilepton production 
~\cite{Shuryak:1978ij} 
in intermediate range of invariant mass (say $1-3$ GeV) with optimized contribution 
from the 
QGP, which is not dominated by hadronic processes, but still treated via perturbative 
methods. We emphasize that the higher order perturbative 
calculations~\cite{Aurenche:1998nw, Laine:2013vma, Ghisoiu:2014mha, 
Ghiglieri:2014kma} 
of the dilepton rate do not converge in a small strong coupling ($g$) limit. This is 
because the
temperatures attained in recent heavy-ion collisions are not so high to make 
perturbative calculations applicable. However, the leading order perturbative 
quark-antiquark annihilation is the only dilepton rate from the
QGP phase that has been used extensively in the literature. Nevertheless, 
this contribution is very appropriate at large invariant mass but not in low and 
intermediate invariant mass. In this mass regime one expects that the nonperturbative 
contributions could be important and substantial. 

The nonperturbative effects of QCD are taken care by the lattice QCD (LQCD) 
computations, a first principle based method of QCD. It 
has very reliably computed the nonperturbative effects associated with the bulk 
properties (thermodynamics and conserved density fluctuations) of the deconfined phase, 
around and above the deconfined temperature. Further, the efforts have also been made in 
lattice within the quenched  approximation of QCD~\cite{Ding:2010ga, 
Ding:2016hua, Karsch:2001uw, Aarts:2002cc, Kaczmarek:2011ht,Aarts:2005hg} and in full 
QCD~\cite{Amato:2013naa, Aarts:2014nba} for studying the structure of vector 
correlation functions and their spectral representations.  Nevertheless, such studies 
have provided \textit{only} 
critically needed information about various transport 
coefficients both at zero~\cite{Aarts:2002cc, Kaczmarek:2011ht} 
and finite~\cite{Aarts:2005hg} momentum, and the thermal 
dilepton rate~\cite{Karsch:2001uw, Ding:2010ga, Ding:2016hua}. The computation of these 
quantities proceed by first evaluating the Euclidean time correlation function only for a 
finite set of discrete  Euclidean times to reconstruct the vector spectral function in 
continuous real time using maximum entropy method (MEM)~\cite{Asakawa:2000tr, 
Nakahara:1999vy}, thereby extracting various 
spectral properties. Unfortunately the lattice techniques are solely applicable in 
Euclidean spacetime, while the spectral function is an inherently Minkowskian object. 
Though it can be obtained from the Euclidean correlator in principle, but the process of 
analytic continuation in the regime of lattice is ill-defined. Because of this 
complication, in LQCD the spectral function is not defined via Eq.~(\ref{spec_def}) but 
through a probabilistic method MEM~\cite{Asakawa:2000tr,Nakahara:1999vy}, which is also 
in some extent error prone~\cite{Cuniberti:2001hm}. Nevertheless, because of its 
limitations LQCD data~\cite{Karsch:2001uw, Ding:2010ga, Ding:2016hua} also could not 
shed much light on the low and intermediate mass dileptons as it is indeed a difficult 
task in lattice.

It is now desirable to have an alternative approach to include nonperturbative effects
in dilepton production. It is well known that the 
QCD vacuum has a nontrivial structure consisting of non-perturbative fluctuations of the 
quark  and gluonic fields. These fluctuations can be traced via a few phenomenological 
quantities, known as vacuum condensates~\cite{GellMann:1968rz}. In standard perturbation 
theory for simplicity one works with an apparent vacuum and the theory becomes less 
effective with relatively lower invariant mass.  The vacuum expectation values of the 
such condensates vanishes in the perturbation theory by definition. But in 
reality they are non-vanishing~\cite{Vainshtein:1978sz, Lavelle:1988eg} and thus the
idea of the nonperturbative dynamics of QCD is signaled by the emergence of power 
corrections in physical observables through the inclusion of nonvanishing vacuum 
expectation values of local quark and gluonic operators such as the quark
and gluon condensates.  In present calculation we intend to 
compute 
intermediate mass (IM) dilepton production using a nonperturbative power corrections.

In this context Shifman-Vainshtein-Zakharov (SVZ) first argued~\cite{Shifman:1978bx, 
Shifman:1978by}  that Wilson's Operator Product Expansion (OPE)~\cite{Wilson:1969zs} is  
valid in presence of the non-perturbative effects~\cite{Novikov:1980uj}. By using OPE 
judiciously one can exploit both perturbative and non-perturbative domain 
separately~\cite{Hubschmid:1982pa, Mallik:1983nn, Bagan:1992tg}. Unlike QED a favorable 
situation occurs particularly in QCD that allows us to do the power 
counting~\cite{Novikov:1984rf,Novikov:1984ac}. OPE basically assumes a separation of 
large and short distance effects via condensates and Wilson coefficients. Also according 
to SVZ, the less effectiveness of ordinary perturbation theory at relatively low 
invariant mass is a 
manifestation of the fact that nonperturbative vacuum condensates are appearing as power 
corrections in the OPE of a Green's function. So in view of OPE, in the large-momentum 
(short-distance) limit,  a two point current-current correlation 
function~\cite{Shifman:1978bx, 
Shifman:1978by, Narison:2002pw} can be 
expanded in terms  of local composite operators  and 
$c$-numbered Wilson coefficients as
\bea
C(p) &=\atop{z \rightarrow 0}& i\int e^{ip\cdot z}d^4z \big \langle {\cal T} 
\left\{J(z) 
J(0)\right\}\big \rangle
= \sum \limits_{n} 
W_n(p^2,\nu^2)\ \langle  O_n \rangle_D, \label{corr_func}
\eea 
provided $p^2 >> \Lambda^2$ , where $\Lambda$  is the QCD scale and 
${\cal T}$ is the time ordered product. $O_n$ is the  $D$-dimensional  composite 
operators (condensates) and  have non-zero expectation values which were absent to all 
orders in perturbation theory. $\nu$ is a factorization scale that separates long and 
short distance 
dynamics. The power corrections appear through the Wilson coefficients $W$ that contain 
all information about large momentum (short distance) physics above the scale $\nu$, 
implying that those are free from any infrared and nonperturbative long distance effects. 
Alternatively, the factorization scale  $\nu$ is chosen to minimize the perturbative 
contributions to 
the condensates such that the physical observable should, in principle, be insensitive to 
the choice of $\nu$. We note that for  
computing a correlator in vacuum(medium) one should first calculate it in a background
of quark and gluonic fields and then average it with respect to these fields 
in the vacuum(medium) to incorporate the power corrections through relevant condensates.

Before going into our calculation we would like note following points in OPE:
The general and important issue in OPE is the separation of various scales. At finite 
temperature
the heatbath introduces a scale $T$, and then OPE  has three scales: $\Lambda$, $T$ and 
$p$ beside the factorization scale\footnote{ One chooses the factorization scale $\nu$ as
$\Lambda\sim T\lesssim \nu \ll p$ above which the state dependent fluctuations reside in 
the expectation values of the operators~\cite{CaronHuot:2009ns}.} 
$\nu$. 
Based on this one can have\footnote{We also note that there can 
be another one: $\Lambda$ soft, $p$ 
hard and $T$ super-hard ($\Lambda < p < < T$). In this case there is a double
scale separation and one does not need it for OPE.} either (i) $\Lambda$ and $T$ soft but 
$p$ hard 
($\Lambda\sim T < p$) or (ii)  $\Lambda$ soft  but $p$ and $T$ hard ($\Lambda < p\sim 
T$). 
\begin{enumerate}
 \item The general belief~\cite{Shuryak:1988ck,Bochkarev:1985ex} that the 
Wilson coefficients 
$W(p^2,\nu^2)$ are c-numbered and remain same 
irrespective of the states considered. This means if one takes vacuum average $\langle 
\cdots \rangle_0$ or thermal average $\langle \cdots \rangle_\beta$ of 
eq.(\ref{corr_func}),  the Wilson coefficient functions, $W(p^2,\nu^2)$, remain 
temperature independent whereas the temperature dependence resides only in $\langle O 
\rangle_\beta$. In other way,  OPE is an expansion in $1/p$ where $p$ is
the typical momentum scale. But thermal effects are essentially down by
thermal factors $\exp(-p/T)$. For $\Lambda\sim T < p$, this does not contribute to any 
order in $1/p$ in OPE. 
It is like $\exp(-1/x)$ for which all coefficients in the Taylor expansion in $x(=T/p)$ 
vanish. 

\item  There are also efforts to extend the OPE to a system with finite 
temperature~\cite{Furnstahl:1989ji,Hansson:1990tt,Bochkarev:1985ex,Bochkarev:1984zx,
Dosch:1988vt}.
At nonzero $T$ the heatbath introduces perturbative 
contributions to the matrix element in OPE, $\langle O \rangle_\beta$ in addition to the 
nonperturbative contributions of finite dimension composite operator. One needs to 
determine a temperature $T\sim p$ above which the perturbative calculation of thermal 
corrections is reliable. Usually these perturbative 
thermal corrections are incorporated by making the Wilson coefficients temperature 
dependent through systematically resummed infinite order in the expansion that comes out 
to be $\sim T$ (but not $\sim gT$). This means if one
takes the thermal average of eq.(\ref{corr_func}), then one requires contributions to 
infinite order in the expansion to get $W$'s temperature dependent, and it
becomes  an expansion of 
$\Lambda/p$ whereas the expansion in $1/(p/T)$ is already 
resummed~\cite{Furnstahl:1989ji}. Nevertheless, this 
resummation is appropriate when $T\gtrsim p$,  but QCD sum rule approach may break down 
and lose its predictive power~\cite{Hansson:1990tt}.  

However,  for low temperature ($\Lambda \sim T< p$) such 
resummation does not make 
much sense. Thus for low  $T$  the temperature acts as an infrared effect and cannot 
change 
the Wilson 
coefficients. Since our calculation  is intended for low temperature ($\pi T< p<\omega$), 
the temperature dependence is only considered in the condensates based on the above point 
1 vis-a-vis scale separtion as in case (i).
\end{enumerate}

Now the in-medium differential dilepton production rate~\cite{Weldon:1990iw, 
McLerran:1984ay} is 
related to the electromagnetic spectral function as 
\bea
\label{dilepton_rate_eqn}
\frac{dR}{d^4xd^4p} =  \frac{\alpha_{\textrm em}}{12 \pi^3 M^2}
n_B \left(\omega\right)  \sum\limits_f q_f^2 \, \rho_f(\omega, |\vec p|),  
\eea
where $n_B(x)=(e^x-1)^{-1}$ is the Bose-Einstein distribution function, $q_f$ is the 
electric charge of a given quark flavor $f$, 
$\alpha_{\textrm em}$ is the electromagnetic fine structure constant and the 
invariant mass of the lepton pair is $M^2 =p^2= \omega^2 - \vert \vec{p}\vert^2$ with 
$p\equiv(\omega, \vec p)$. The dilepton rate in (\ref{dilepton_rate_eqn}) is valid only 
at leading order in $\alpha_{\textrm em}$ but to all orders in strong coupling constant 
$\alpha_s$. The quark and lepton masses are neglected in 
(\ref{dilepton_rate_eqn}).

The electromagnetic spectral function for a given flavor $f$, $\rho_f\left(\omega,|\vec 
p| \right)$, is extracted from the 
timelike discontinuity of the two point correlation function as
\bea  
\rho_f\left(\omega,|\vec p| \right) 
= \frac{1}{\pi}\, \mathrm{Im} 
\left (C^{\mu}_{\mu}\right )_f.
\label{spec_def}
\eea   
The main aim 
of the present paper is to obtain the in-medium electromagnetic spectral function 
incorporating the power corrections within OPE in $D=4$ dimension and analyze its effect 
on the thermal dilepton rate from QGP. To obtain the in-medium electromagnetic spectral 
function one needs to calculate the two point correlation function  via OPE corresponding 
to the $D=4$ gluonic and quark operators (condensates) in hot QCD medium. The power 
corrections appears in the spectral function through the nonanalytic behavior of the 
correlation function in powers of $p^{-D/2}$ or logarithms in the Wilson coefficients 
within OPE in $D$ dimension.

The plan of the paper is as follows. In section \ref{setup} we outline some generalities 
needed for the purpose. In sections \ref{quark_op} and 
\ref{gluon_op} we discuss how in-medium quark and gluonic composite operators 
in $D=4$, respectively, can be included in electromagnetic polarization diagram.
We then obtain the two point correlators in terms of Wilson coefficients and   
those composite operators for the case of light quarks. 
We also demonstrate how the mass singularity appearing in the correlator is absorbed by 
using minimal subtraction via operator 
mixing. In section 
\ref{spectral}  we discuss about the thermal spectral function and it's modification due 
to incorporation of leading order power correction, particularly in the range of 
intermediate invariant mass.
As a spectral property the dilepton production is discussed  in section 
\ref{dilepton_rate} and compare our results with some other known perturbative and 
nonperturbative results and then we conclude in section \ref{conclusion}.

\section{Setup}
\label{setup}
\begin{center}
\begin{figure}[h]
 \begin{center}
 \includegraphics[scale=0.4]{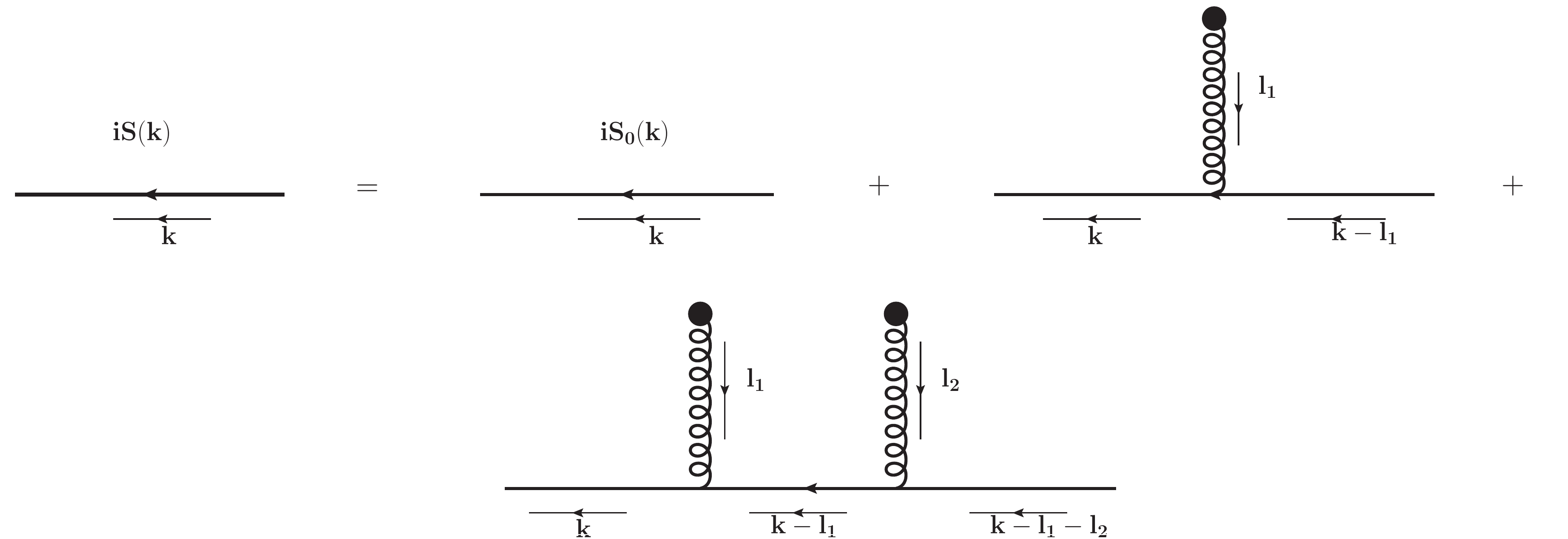} 
 \caption{Effective quark propagator in the background gluon field.}
  \label{quark_propag}
 \end{center}
\end{figure}
\end{center}

We briefly outline some generalities which are essential ingredients in our calculation. 
In usual notation the nonabelian field tensor in $SU(3)$ is defined as
\bea
G_{\mu\nu}^a &=& \partial_\mu A_\nu^a - \partial_\nu A_\mu^a + 
gf^{abc}A_\mu^bA_\nu^c,\nn\\
 G_{\mu\nu}&=&G_{\mu\nu}^at^a, \nn\\
D_\alpha &=& \partial_\alpha-igt^aA_\mu^a,\
% \langle G_{\mu\nu}^a(0) G_{\alpha\beta}^b(0) \rangle &=& \frac{1}{96}\delta^{ab}
% (g_{\mu\alpha}g_{\nu\beta}-g_{\mu\beta}g_{\nu\alpha})\langle G_{\rho\sigma}^c 
% G^{c~\rho\sigma} \rangle.
\label{vacuum_fields}
\eea
where $a,b,c$ are color indices, $t^a$ are the generators and 
$g_{\mu\nu}=\text{diag}(1,-1,-1,-1)$. Consequently in vacuum it satisfies the projection 
relation for composite operator
\bea
\Big\langle G_{\mu\nu}^a(0) G_{\alpha\beta}^b(0)\Big \rangle &=& \frac{1}{96}\delta^{ab}
(g_{\mu\alpha}g_{\nu\beta}-g_{\mu\beta}g_{\nu\alpha})\Big \langle G_{\rho\sigma}^c 
G^{c~\rho\sigma} \Big \rangle.
\label{vacuum_projection}
\eea
Choosing the Fock-Schwinger aka the fixed point gauge ($x^\mu A_\mu^a(x)=0$) for 
convenience, the gauge field $A_\mu^a(x)$ can be expressed easily in terms of 
gauge covariant quantities~\cite{Smilga:1982wx, 
Reinders:1984sr,Novikov:1984rf,Novikov:1984ac} as 
\bea
A_\mu(x)=\int\limits_0^1 \sigma ~ d\sigma ~G_{\nu\mu}(\sigma x)x^\nu =\frac{1}{2} 
x^\nu G_{\nu\mu}(0)+\frac{1}{3}x^\alpha x^\nu D_\alpha G_{\nu\mu}(0)+\cdots \, , \nn
\eea
where first the gauge field $G(y)$ has been Taylor expanded in the small $\sigma$ limit 
and then the integration over $\sigma$ has been performed. Now in momentum space it reads
\bea
A_\mu(k)&=& \int A_\mu(x) e^{ikx} d^4x\nn\\
&=&\frac{-i(2\pi)^4}{2}G_{\nu\mu}(0)\frac{\partial}{\partial k_\nu}\delta^4(k)+
\frac{(-i)^2(2\pi)^4}{2}(D_\alpha G_{\nu\mu}(0))\frac{\partial^2}{\partial k_\nu \partial 
k_\alpha}\delta^4(k)+\cdots,
\label{gauge_field}
\eea
where each background gluon line will be associated with a momentum integration as we 
will see below.
Using this one can now evaluate the effective quark propagator in presence of 
background gluon lines~\cite{Novikov:1984rf,Novikov:1984ac} by 
expanding the number of gluon legs attached to the bare quark line as in 
Fig.~\ref{quark_propag}. So, it can be written as
\bea
S_{\textrm{eff}}&=& S_0 + S_1 + S_2 + \cdots , \label{quark_propagator_eff} 
\eea
where the bare propagator for massive quark reads as
\bea
S_0&=&\frac{i}{\slashed{k}-m} \label{quark_propagator1}
\eea
where $m$ is the mass of the quark.

With one gluon leg attached to the bare quark (Fig.~\ref{quark_propag}) the expression 
reads as
\bea
S_1&=& \frac{-i}{\slashed{k}-m}\int 
\frac{d^4l_1}{(2\pi)^4}\frac{\slashed{A}(l_1)}{\slashed{k}-\slashed{l_1}-m}\nn\\
&=&-\frac{i}{4}gt^aG_{\mu\nu}^a(0)\frac{1}{(k^2-m^2)^2}\{\sigma^{\mu\nu}(\slashed{k}
+m)+(\slashed{k}+m)\sigma^{\mu\nu}\} \label{quark_propagator2},
\eea
where 
\bea
\sigma^{\mu\nu} &=& \frac{i}{2}[\gamma^\mu,\gamma^\nu], \nonumber
\eea
and the background gauge field $\slashed{A}(l_1)$ is replaced by the first term of 
the gauge field as given in Eq.~(\ref{gauge_field}).
Similarly for the diagram where two gluon legs are attached to the bare quark, we get
\bea
S_2 &=& \frac{i}{\slashed{k}-m}\int 
\frac{d^4l_1}{(2\pi)^4}\frac{\slashed{A}(l_1)}{\slashed{k}-\slashed{l_1}-m}\int 
\frac{d^4l_2}{(2\pi)^4}\frac{\slashed{A}(l_2)}{\slashed{k}-\slashed{l_1}-\slashed{l_2}-m}
\nn\\
&=&-\frac{i}{4}g^2t^at^bG_{\alpha\beta}^a(0)G_{\mu\nu}^b(0)\frac{(\slashed{k}+m)}{
(k^2-m^2)^5}(f^{\alpha\beta\mu\nu}+f^{\alpha\mu\beta\nu}+f^{\alpha\mu\nu\beta}), 
\label{quark_propagator3}
\eea
where
\bea
f^{\alpha\beta\mu\nu} &=& \gamma^\alpha(\slashed{k}+m)\gamma^\beta(\slashed{k}+m) 
\gamma^\mu(\slashed{k}+m)\gamma^\nu(\slashed{k}+m).\nonumber
\eea
In presence of a medium, however, a four-vector $u_\mu = (1,0,0,0)$ is usually introduced 
to restore Lorentz invariance in the rest frame of 
the heat bath. So, at finite 
temperature additional scalar operators can be constructed so that the vacuum operators 
are generalized to in-medium ones. The projection relation of composite operator in 
(\ref{vacuum_projection}) gets modified in finite temperature~\cite{Mallik:1997pq, 
Mallik:1983nn, Antonov:2004mt, 
Basar:2014swa} as 
\bea
\big \langle G_{\mu\nu}^a(0) G_{\alpha\beta}^b(0) \big \rangle_T &=& 
\big [g_{\mu\alpha}g_{\nu\beta}-g_{\mu\beta}g_{\nu\alpha}\big ]A  
 - \big [(u_\mu u_\alpha g_{\nu\beta}-u_\mu u_\beta g_{\nu\alpha}-u_\nu u_\alpha 
g_{\mu\beta}  \nn \\
&&  +
u_\nu u_\beta 
g_{\mu\alpha})-\frac{1}{2}(g_{\mu\alpha}g_{\nu\beta}-g_{\mu\beta}g_{\nu\alpha})\big ]B 
 + i\epsilon_{\mu\nu\alpha\beta}C ,
\label{medium_projection}
\eea
where $A,B \, \textrm{and} \, C$ are, respectively,  given as
\bea
A = \frac{\delta^{ab}}{96}\big \langle G^2 \big \rangle_T,~ B = 
\frac{\delta^{ab}}{12}\big \langle u\Theta^gu \big \rangle_T,~ C = 
\frac{\delta^{ab}}{96}\big \langle {\cal E} \cdot {\cal B} \big \rangle_T \nn
\eea
with ${\cal E}$ and ${\cal B}$ are, respectively, the electric and magnetic fileds.  
The traceless gluonic  stress tensor, $\Theta^g_{\mu\nu}$,  is given by 
\bea
\Theta^g_{\mu\nu} = 
-G_{\mu\rho}^aG_\nu^{a\rho}+\frac{1}{4}g_{\mu\nu}G^a_{\rho\sigma}G^{\rho\sigma~a}.
\eea
QCD vacuum consists of both quark and gluonic fields. We note that the composite 
operators involving quark fields will be defined below whenever necessary. 

\section{Composite Quark Operators}
\label{quark_op}

\begin{center}
\begin{figure}[h]
 \begin{center}
 \includegraphics[scale=0.3]{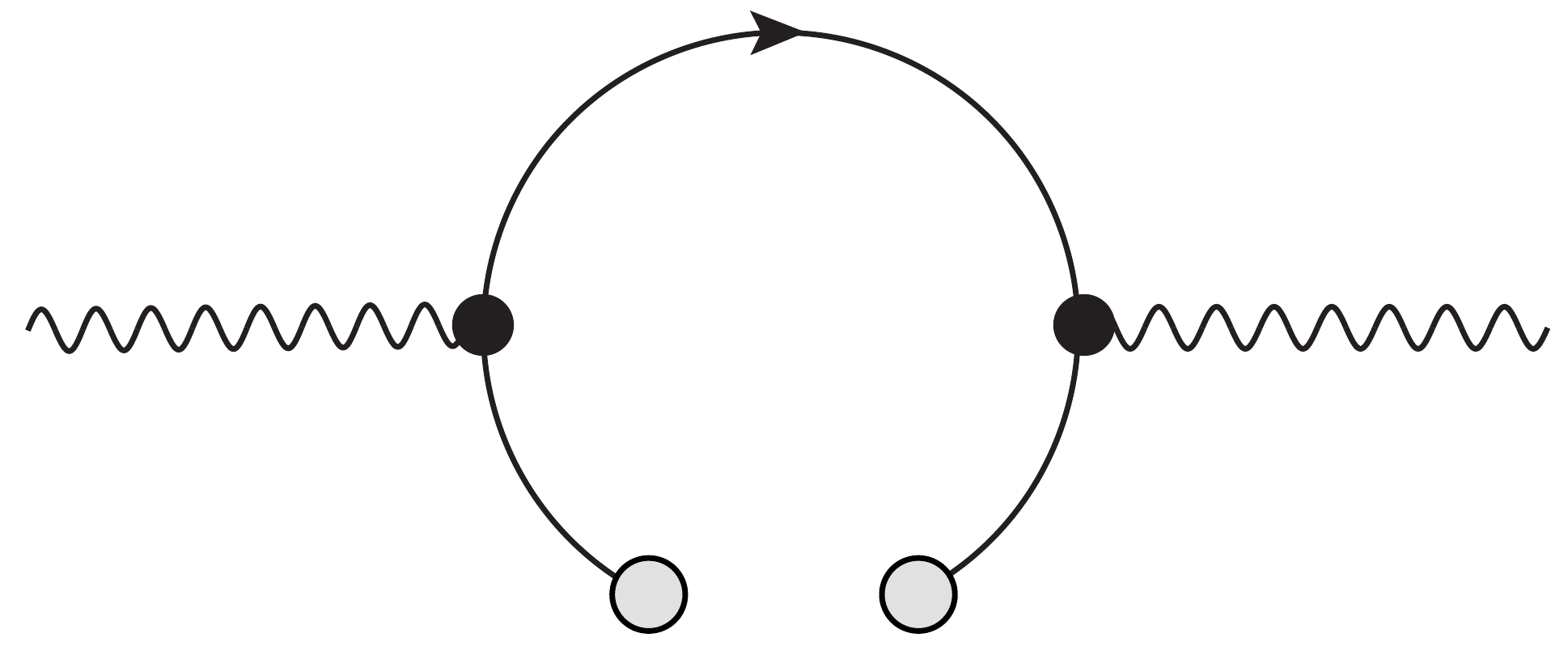}\hspace*{0.5cm} 
 \includegraphics[scale=0.3]{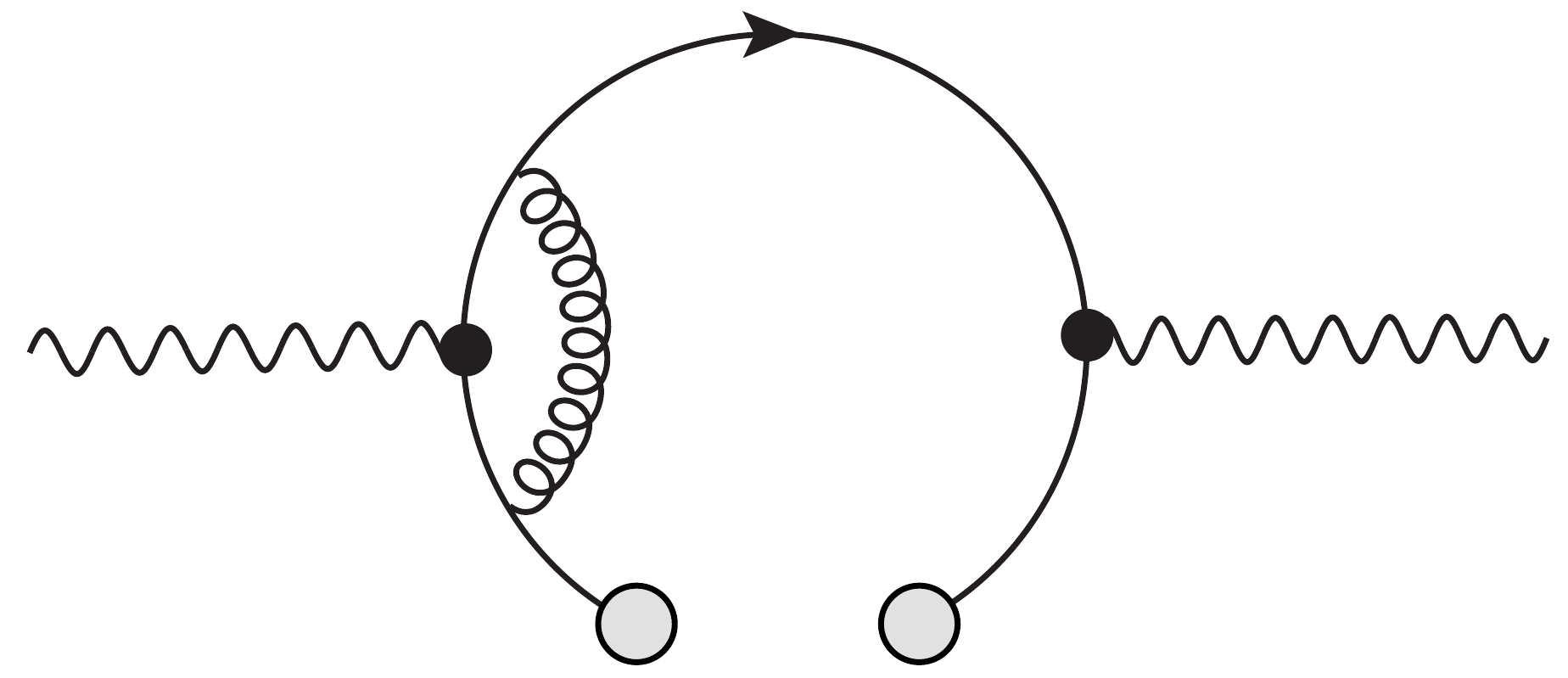} \\
 \vspace*{0.6cm}
 \includegraphics[scale=0.3]{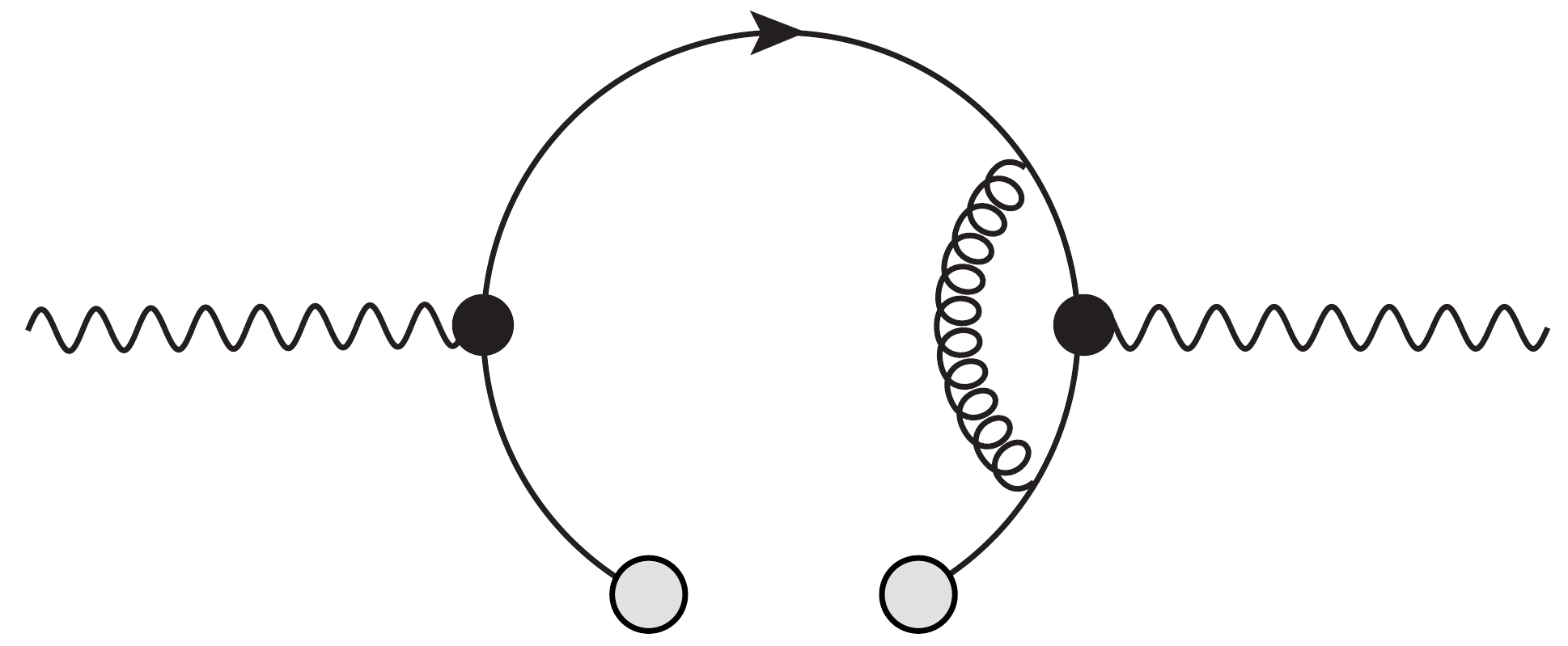}\hspace*{0.5cm} 
 \includegraphics[scale=0.3]{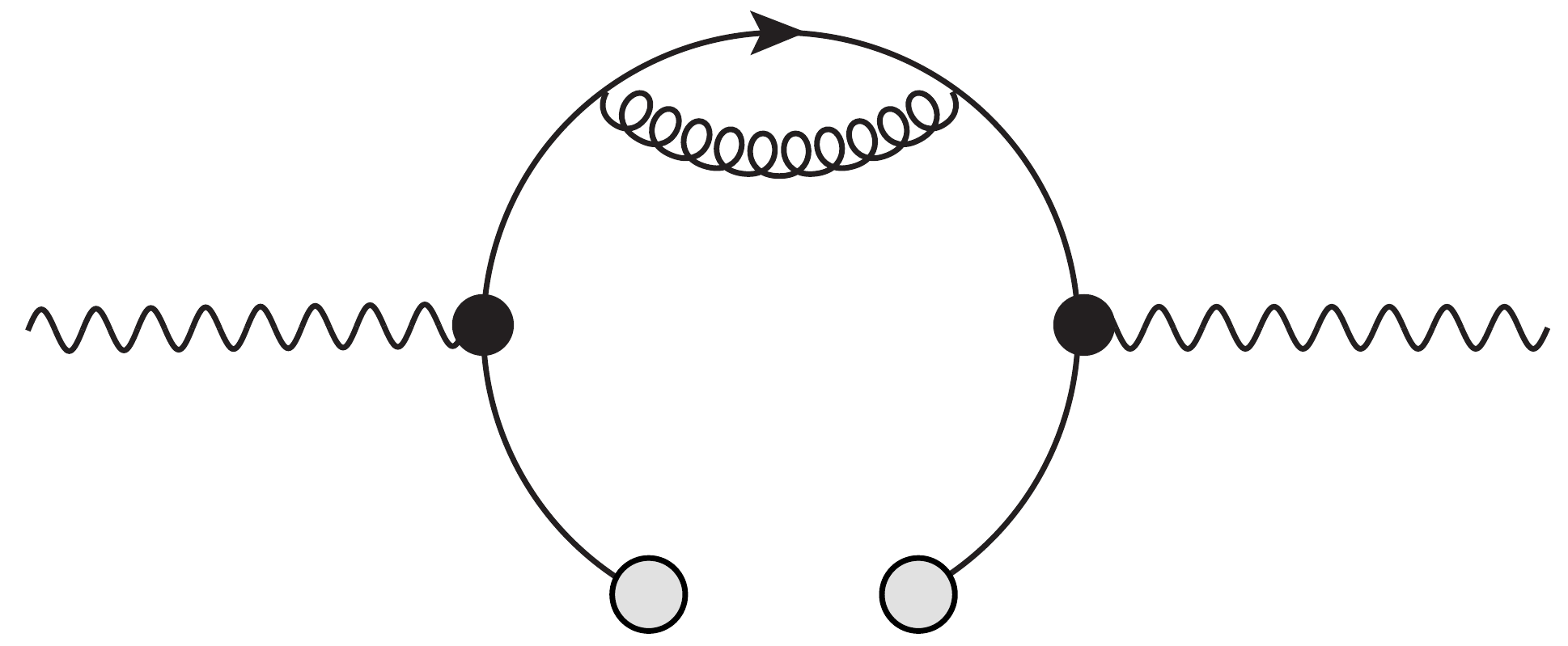} 
 \caption{OPE topologies corresponding to  $D=4$ composite quark operator that can 
contribute to the power correction in the one-loop electromagnetic polarization. The 
composite quark operator  
is denoted by two gray blobs separated by a gap in the internal quark line. Top left one 
contributes to the leading order (LO:$e^2$) whereas the remaining three are due to  
gluonic corrections contribute to non-leading order (NLO: $e^2g^2$).}
  \label{quark_condensate}
 \end{center}
\end{figure}
\end{center}

In this section we would like to discuss  the power 
corrections using  composite quark operators (condensates) in OPE. In presence of quark 
condensate the leading order contribution comes from the top left 
panel in Fig.~\ref{quark_condensate} where one of the internal quark line  is soft in 
the polarization diagram  represented by two gray blobs with a gap. We note that the 
diagrams here are not thermal field theory diagrams but$``$OPE diagrams" 
with scales: $\Lambda$ and $T$ soft but $p$ hard ($\Lambda \sim T< p$).
The 
corresponding contributions can be obtained as
\bea
\Big[C^{\textrm{LO}}_{\mu\nu}(p)\Big ]_q &=& i\int e^{ip\cdot z}d^4z \big \langle 
{\cal T}\{j_\mu(z)j_\mu(0)\}\big \rangle\nn\\
&=& -N_cN_f\int e^{ip\cdot z}d^4z \textrm{Tr}\big[\bar{\psi}(z)\gamma_\mu S(z,0)  
\gamma_\nu \psi(0) + \bar{\psi}(0)\gamma_\nu S(0,z) \gamma_\mu \psi(z)\big], \label{lo_Q}
\eea
where  $N_f$ is the number of quark flavor and $N_c$ is the number of color for a 
given flavor. We also note that the soft quark lines are represented by Heisenberg 
operators $\psi(z)$ and $\psi(0)$. In the large $p$ limit, $\psi(z)$ can be expanded as
\bea
\psi(z)=\psi(0)+ z^\mu D_\mu \psi(0).\label{exp_heisen}
\eea
Now considering only the first term in the expansion of (\ref{exp_heisen}) 
in (\ref{lo_Q}), one gets
\bea
\Big[C^{\textrm{LO}}_{\mu\nu}(p)\Big]^1_q &=& -N_cN_f\frac{1}{12}\big \langle 
\bar{\psi}\psi\big \rangle\textrm{Tr}\big[\gamma_\mu S(p) \gamma_\nu + \gamma_\nu S(p) 
\gamma_\mu\big ], \label{lo_Q1}
\eea
which, as expected, vanishes in the chiral limit. This is because of the 
appearance of the chiral condensate, $\langle \bar{\psi}\psi\rangle$  which 
is proportional to  the quark mass $m$. On the other hand, choosing the second term in 
the expansion of (\ref{exp_heisen}) we get 
\bea
\Big[C^{\textrm{LO}}_{\mu\nu}(p)\Big]^2_q\!\! &=&\!\! -N_cN_f \!\! \int e^{ip\cdot z} \, 
\, d^4z \, \, \times\nn\\ 
&&\textrm{Tr}\left[\bar{\psi}(0)z^\rho \overleftarrow{D}_\rho\gamma_\mu S(z,0) \gamma_\nu 
\psi(0) + \bar{\psi}(0)\gamma_\nu S(0,z) \gamma_\mu z^\rho 
\overrightarrow{D}_\rho\psi(0)\right].\label{lo_Q2}
\eea
Now the most general decomposition of $\langle \bar{\psi}iD_\rho \psi \rangle$ 
for the massless at finite temperature~\cite{Mallik:1997pq} is  given  as
\bea
\langle \bar{\psi}iD_\rho \psi \rangle_T &=& \left(-\frac{1}{12} 
\gamma_\rho+\frac{1}{3}u_\rho \slashed{u}\right) \langle  u\Theta^fu \rangle_T, 
\label{quark_ope} 
\eea
where $\Theta^f_{\mu\nu}$ is traceless fermionic stress tensor and in the 
massless limit it is given by
\bea
\Theta^f_{\mu\nu} = \bar{\psi}\gamma_\mu i D_\nu \psi.
\label{quark_condensate_relation1}
\eea
Using (\ref{quark_ope}) in (\ref{lo_Q2}), and performing $z$-integration  one gets
\bea
\Big[C^{\textrm{LO}}_{\mu\nu}(p)\Big]^2_q &=& -N_cN_f~\frac{\partial}{\partial 
p^\rho}\textrm{Tr}\Big[\big(\gamma_\mu S(p) \gamma_\nu+\gamma_\nu S(p) 
\gamma_\mu\big)\big(-\gamma_\rho+4u_\rho \slashed{u} \big)\Big]\frac{1}{12}
\big \langle u\Theta^fu \big \rangle_T. \label{lo_Q2_c}
\eea
Now treating the Wilson coefficients temperature independent, the LO 
contribution is obtained as
\bea
\Big[C^{\mu \, \textrm{LO}}_\mu(p)\Big]_q &=& \Big[C^{\mu \, \textrm{LO}}_\mu(p)\Big]^2_Q 
= 
\frac{8N_cN_f}{3p^2}\left(1-4\frac{\omega^2}{p^2}\right)\big\langle u\Theta^fu 
\big\rangle_T. \label{lo_q2_final} 
\eea
We note that the contributions from NLO order gluonic corrections (Fig. 
\ref{quark_condensate}) to  quark vacuum condensates are already evaluated 
in~\cite{Pascual:1981jr}. Following the same prescription as LO   
the total NLO in-medium contributions from remaining three diagrams in 
Fig.\ref{quark_condensate} in the massless limit is obtained as 
\bea
\Big[C^{\mu \, \textrm{NLO}}_\mu(p)\Big]_q &=& \frac{8N_cN_f}{3p^2}\big\langle u\Theta^fu 
\big\rangle_T\left(1-4\frac{\omega^2}{p^2}\right)
\frac{2g^2}{9\pi^2}\left(1-\ln\left(\frac{-p^2}{\Lambda^2}\right)\right).
\label{nlo_Q}
\eea
We note that the logarithmic correction appears from the radiative correction diagrams 
in Fig.\ref{quark_condensate} when the ultraviolet divergences associated with them are 
regularized through dimensional regularization~\cite{peskin}. The non-analytic behaviour 
of this logarithmic term will generate the power tail in the spectral function.

Combining (\ref{lo_q2_final}) and (\ref{nlo_Q}) one obtains  power 
correction upto NLO due to quark operator in the electromagnetic correlation function at 
finite temperature as 
\bea
\Big [C_\mu^\mu(p) \Big ]_q &=& \frac{8N_cN_f}{3p^2}\big\langle u\Theta^fu 
\big\rangle_T\left(1-4\frac{\omega^2}{p^2}\right)
\left[1+\frac{2g^2}{9\pi^2}\left(1-\ln\left(\frac{-p^2}{\Lambda^2}\right)\right)\right].
\label{quark_operator_massless}
\eea

\section{Composite Gluonic Operators}
\label{gluon_op}
\begin{center}
\begin{figure}[h]
 \begin{center}
 \includegraphics[scale=0.5]{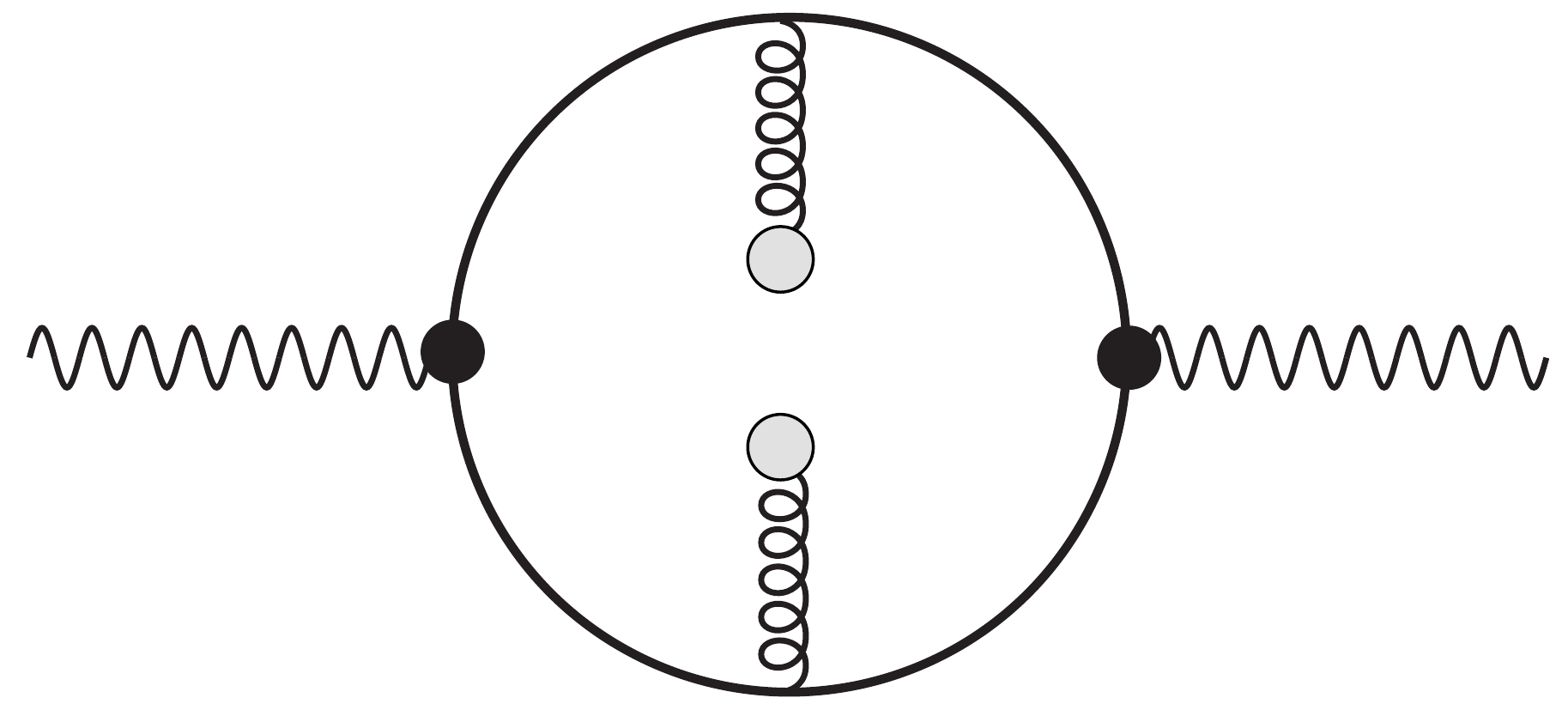} 
 \caption{(Topology-I) Vertex correction  where one soft gluon line is attached to 
each internal quark line  in the electromagnetic polarization diagram.}
  \label{topology_1}
 \end{center}
\end{figure}
\end{center}

\begin{center}
\begin{figure}[h]
 \begin{center}
\includegraphics[scale=0.5]{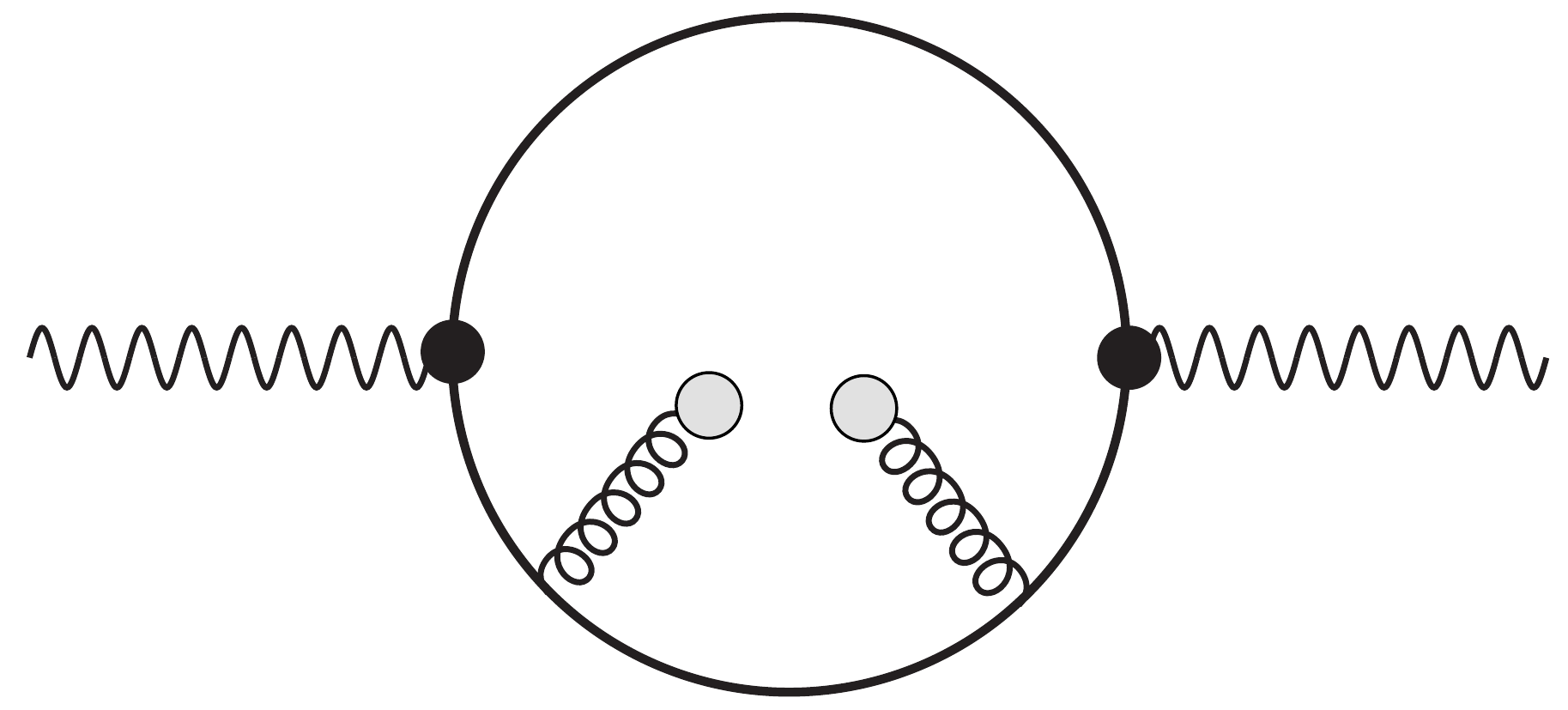}
\caption{(Topology-II) Self-energy correction where two soft gluon lines are attached to 
one internal quark line in the electromagnetic polarization diagram. A similar topology 
will also arise when two soft gluon lines are attached to the other quark line.}
\label{topology_2}
\end{center}
\end{figure}
\end{center}

In this section we compute the power correction to the electromagnetic correlation 
function  from the $D=4$ composite gluonic operator by considering the soft  
gluon 
lines attached to the internal quark lines in the electromagnetic polarization diagram. 
There are two such topologies, as shown in Figs. \ref{topology_1} and \ref{topology_2}, 
depending upon how the soft gluon line is attached to the internal quark lines in 
electromagnetic polarization diagram.

Using Eqs.(\ref{quark_propagator1})-(\ref{quark_propagator3}),
the contribution of the vertex correction diagram (topology-I) in vacuum can be written 
as 
\bea
\Big[C_{\mu}^{\mu}(p)\Big]_{g}^{\textrm{I}} &=& iN_cN_f\int 
\frac{d^4k}{(2\pi)^4}  
\textrm{Tr} \Big [ \gamma_\mu S_1(k)\gamma^\mu S_1(q)\Big ] \nn\\
&=& -\frac{iN_cN_f}{16}g^2t^at^b\Big\langle G_{\rho\sigma}^a(0) G_{\alpha\beta}^b(0) 
\Big\rangle \nn\\
&&\times \int \frac{d^4k}{(2\pi)^4} \frac{\textrm{Tr} \Big[\gamma_\mu 
\left(\sigma^{\rho\sigma} 
(\slashed{k}+m)+(\slashed{k}+m)\sigma^{\rho\sigma}\right)\gamma^\mu 
\left(\sigma^{\alpha\beta}(\slashed{q}+m)+(\slashed{q}+m)\sigma^{\alpha\beta}\right)\Big 
]} { (k^2-m^2)^2(q^2-m^2)^2}\nn\\
&=& 
-iN_cN_f\Big\langle g^2G^2 \Big\rangle \int \frac{d^4k}{(2\pi)^4} \frac{k\cdot q} 
{(k^2-m^2)^2(q^2-m^2)^2}, \label{topologyI_vacuum}
\eea
where $q=k-p$  and  we have also used Eq.(\ref{vacuum_projection}) in the last step after 
evaluating the trace.

Similarly, the contribution from the topology-II (including 
the one where two gluon lines are attached to the other quark propagator in 
Fig.\ref{topology_2}) in vacuum can be written as
\bea
\Big[C_{\mu}^{\mu}(p)\Big]_{g}^{\textrm{II}} &=&  2 iN_cN_f\int \frac{d^4k}{(2\pi)^4} 
\textrm{Tr} \Big[\gamma_\mu S_2(k)\gamma^\mu S_0(q)\Big]\nn\\
&=& \frac{iN_cN_f}{4}g^2t^at^b\Big\langle G_{\rho\sigma}^a(0) G_{\alpha\beta}^b(0) 
\Big\rangle \nn\\
&&\times \int \frac{d^4k}{(2\pi)^4} \frac{\textrm{Tr} \Big[ \gamma_\mu 
(\slashed{k}+m)(f^{\alpha\beta\mu\nu}
+f^{\alpha\mu\beta\nu}+f^{\alpha\mu\nu\beta})\gamma^\mu 
\left(\slashed{q}+m\right)\Big]}{(k^2-m^2)^5(q^2-m^2)}\nn\\
 &=& 
-iN_cN_f\Big\langle g^2G^2 \Big\rangle \int \frac{d^4k}{(2\pi)^4} 
\frac{4m^2 (k\cdot q-2k^2)}{(k^2-m^2)^4(q^2-m^2)}.\label{topologyII_vacuum}
\eea

Now, we would like to compute both topologies in presence of $D=4$ composite gluonic 
operators at finite $T$. For the purpose,  unlike vacuum case one requires to use 
in-medium gluon condensates as given in Eq.(\ref{medium_projection}). The
vacuum contributions in Eqs.(\ref{topologyI_vacuum}) and (\ref{topologyII_vacuum}) are, 
respectively, modified at finite $T$ as
\bea
\Big[C_{\mu}^{\mu}(p)\Big]_{g,T}^{\textrm{I}}
&=& -\frac{iN_cN_f}{16}g^2t^at^b\Big\langle G_{\rho\sigma}^a(0) G_{\alpha\beta}^b(0) 
\Big\rangle_T \nn\\
&&\times\int \frac{d^4k}{(2\pi)^4} \frac{\textrm{Tr} \Big[\gamma_\mu 
\big(\sigma^{\rho\sigma}(\slashed{k}+m)+(\slashed{k}+m)\sigma^{\rho\sigma}
\big)\gamma^\mu 
\big(\sigma^{\alpha\beta}(\slashed{q}+m)+(\slashed{q}+m)\sigma^{\alpha\beta}\big)\Big]}
{ (k^2-m^2)^2(q^2-m^2)^2}\nn\\
&=& -iN_cN_f\Big\langle g^2G^2 \Big\rangle_T 
\int \frac{d^4k}{(2\pi)^4} \frac{k\cdot q}{(k^2-m^2)^2(q^2-m^2)^2}\nn\\
&& +\frac{4iN_cN_f}{3}\Big\langle g^2 u\Theta^gu \Big\rangle_T 
\int \frac{d^4k}{(2\pi)^4} \frac{(k\cdot 
q-4k_0q_0)}{(k^2-m^2)^2(q^2-m^2)^2}, \label{topologyI_medium}
\eea
and 
\bea
\Big[C_{\mu}^{\mu}(p)\Big]_{g,T}^{\textrm{II}}
&=& \frac{iN_cN_f}{4}g^2t^at^b\Big\langle G_{\rho\sigma}^a(0) G_{\alpha\beta}^b(0) 
\Big\rangle_T \nn\\
&&\times\int \frac{d^4k}{(2\pi)^4} \frac{\textrm{Tr} \Big[\gamma_\mu 
(\slashed{k}+m)(f^{\alpha\beta\mu\nu}+f^{\alpha\mu\beta\nu}+f^{\alpha\mu\nu\beta}
)\gamma^\mu \left(\slashed{q}+m\right)\Big]}{(k^2-m^2)^5(q^2-m^2)}\nn
\eea
\bea
&=& -4iN_cN_fm^2\Big\langle g^2G^2 \Big\rangle_T 
 \int \frac{d^4k}{(2\pi)^4} \frac{(k\cdot 
q-2k^2)}{(k^2-m^2)^4(q^2-m^2)}+\frac{32iN_cN_f}{3}\Big\langle g^2 u\Theta^gu 
\Big\rangle_T\nn\\
&&\times
\int \frac{d^4k}{(2\pi)^4} \frac{k\cdot 
q(2k_0^2-\frac{1}{2}m^2)+m^2(k^2-4k_0^2)-2k_0q_0(k^2-m^2)}{(k^2-m^2)^4(q^2-m^2)}.
\label{topologyII_medium}
\eea
Now, in the short-distance or large-momentum limit of nonperturbative power correction, 
one can work with massless quarks without loss of generality. In the massless limit  
Eq.(\ref{topologyI_medium}) reduces to
\bea
\Big[C_{\mu}^{\mu}(p)\Big]_{g,T}^{\textrm{I}} &=\atop{m\rightarrow 0}& 
-iN_cN_f\Big\langle g^2G^2 \Big\rangle_T 
\int \frac{d^4k}{(2\pi)^4} \frac{k\cdot q}{(k^2)^2(q^2)^2}\nn\\ 
&&+\frac{4iN_cN_f}{3}\Big\langle g^2 u\Theta^gu \Big\rangle_T 
\int \frac{d^4k}{(2\pi)^4} \frac{(k\cdot q-4k_0q_0)}{(k^2)^2(q^2)^2}, 
\label{topologyI_medium_1}
\eea
whereas for Eq.(\ref{topologyII_medium}) the coefficient for $\langle g^2G^2 \rangle_T$ 
vanishes and it takes a simple form 
\bea
\Big[C_{\mu}^{\mu}(p)\Big]_{g,T}^{\textrm{II}} &=\atop{m\rightarrow 0}& 
\frac{32iN_cN_f}{3}\Big\langle g^2 u\Theta^gu \Big\rangle_T 
\int \frac{d^4k}{(2\pi)^4} \frac{2(k\cdot 
qk_0^2-k_0q_0k^2)}{(k^2)^4(q^2)}.\label{topologyII_medium_1}
\eea
Here we emphasize the fact that the vacuum correlation function corresponding to 
topology-II (self-energy correction) in Eq.(\ref{topologyII_vacuum}) vanishes in the 
massless limit. But in medium, one obtains a finite contribution in the massless limit  
as found in Eq.(\ref{topologyII_medium_1}) due to the in-medium 
condensates. Now, the integrals in the above expressions can 
be expressed in terms of standard Feynman integrals $\mathcal{I}_{mn}$, 
$\mathcal{I}_{mn}^\mu$ and $\mathcal{I}_{mn}^{\mu\nu}$ which  have been evaluated in 
appendix \ref{massless_feynman}. Using those results in appendix \ref{massless_feynman}, 
Eqs.(\ref{topologyI_medium_1}) and (\ref{topologyII_medium_1}) in the 
massless limit ($p_0=\omega$), respectively, become
\bea
\Big [C_{\mu}^{\mu}(p)\Big ]_{g,T}^{\textrm{I}} &=\atop{m\rightarrow 0}& -iN_cN_f\langle 
g^2G^2 \rangle_T ~
\frac{1}{2}(2\mathcal{I}_{12}-p^2\mathcal{I}_{22})\nn\\
&& +\frac{4iN_cN_f}{3}\langle g^2 u\Theta^gu \rangle_T~
\frac{1}{2}(2\mathcal{I}_{12}-p^2\mathcal{I}_{22}+8p_0\mathcal{I}_{22}^0-8\mathcal{I}_{22}
^{00})\nn\\
&=&  \langle G^2 \rangle_T~ \frac{g^2N_cN_f}{16\pi^2 p^2} -\langle u\Theta^gu 
\rangle_T~\frac{g^2N_cN_f}{3\pi^2 p^2}
\left[\frac{\omega^2}{p^2}-\frac{1}{4}\right], \label{topologyI_massless} \\
\textrm{\ \ }\ \ \ \ \ \ \ \ \ \ \ \ \ \ && \ \ \ \ \ \ \ \ \ \ \ \ \ \ \ \ \ \ \ \ \ \ \ 
\ \ \ \ \ \ \ \ \ \ \ \ \ \ \  \ \ \ \ \ \nn  \\ 
\textrm{and}\ \ \ \ \ \ \ \ \ \ \ \ \ \ \ \ \ \ \ \ \ \ \  && \ \ \ \ \ \ \ \ \ \ \ \ \ \ 
\ \ \ \ \ \ \ \ \ 
\ \ \ \ \ \ \ \ \ \ \ \ \ \ \  \ \ \ \ \ \nn  \\ 
\textrm{\ \ }\ \ \ \ \ \ \ \ \ \ \ \ \ \ && \ \ \ \ \ \ \ \ \ \ \ \ \ \ \ \ \ \ \ \ \ \ \ 
\ \ \ \ \ \ \ \ \ \ \ \ \ \ \  \ \ \ \ \ \nn  \\ 
\Big[C_{\mu}^{\mu}(p)\Big]_{g,T}^{\textrm{II}} &=\atop{m\rightarrow 0}& 
 \frac{32iN_cN_f}{3}\langle g^2 u\Theta^gu \rangle_T~ 
(2p_0\mathcal{I}_{31}^0-\mathcal{I}_{31}^{00}-p^2\mathcal{I}_{41}^{00})\nn\\
&=& -\frac{g^2N_cN_f}{9\pi^2p^2}\langle u\Theta^gu \rangle_T~
\left[\frac{1}{\tilde{\epsilon}}\left(1-\frac{4\omega^2}{p^2}\right)+2-6\frac{\omega^2}{
p^2}\right].\label{topologyII_massless}
\eea
We note that Eq.(\ref{topologyII_massless}) has a mass singularity as 
${1}/{\tilde{\epsilon}} = 
{1}/{\epsilon}-\ln\left({-p^2}/{\Lambda^2}\right)$ and the reason for which could be 
understood in the following way:  while computing the self-energy correction 
corresponding to topology-II in Fig.~\ref{topology_2}, one actually overcounts a 
contribution from quark condensate. This is because the quark line 
in-between two soft gluon lines in Fig.~\ref{topology_2} becomes soft, leading to 
quark condensate. So the actual contribution from the gluonic operators can only be 
obtained after minimally subtracting the quark condensate 
contribution~\cite{Generalis:1983hb,Broadhurst:1984rr}
which should cancel the mass singularity arising in the massless 
limit\cite{Nikolaev:1982rq, Nikolaev:1982ra, Nikolaev:1981ff}.

To demonstrate this we begin by considering finite quark mass in which a correlator 
containing  quark condensates ($\mathcal{Q}_k$) can be expressed via gluon condensates 
($\mathcal{G}_n$)~\cite{Grozin:1994hd} in mixed representation as
\bea
\mathcal{Q}_k = \sum_n c_{kn}(m)\mathcal{G}_n, \label{mixed_rep}
\eea
where, $c_{kn}(m)$ is an expansion in $1/m$. Then one can also represent a correlator 
with gluon condensates ($\mathcal{G}_n$) as 
\bea
\Big[C_{\mu}^{\mu}(p)\Big]_g &=& \sum_n a_n(p^2,m)\mathcal{G}_n, \label{gc_basis}
\eea
whereas for quark condensates ($\mathcal{Q}_k$) it can be written as
\bea
\Big[C_{\mu}^{\mu}(p)\Big]_q &=& \sum_k b_k(p^2,m)\mathcal{Q}_k, \label{qc_basis}
\eea
with $a_n$ and $b_k$ are the corresponding coefficients for the gluon and quark 
condensates, respectively. Now, in general a correlator with  minimal subtraction
using Eqs.~(\ref{mixed_rep}), (\ref{gc_basis}) and (\ref{qc_basis}) can now be written as 
\bea
\Big[C_{\mu}^{\mu}(p)\Big]_{g}^{\textrm{a}} &=& 
\Big[C_{\mu}^{\mu}(p)\Big]_{g} - \Big[C_{\mu}^{\mu}(p)\Big]_{q} 
= \sum_n a_n(p^2,m)\mathcal{G}_n - \sum_{n,k} b_k(p^2,m)c_{kn}(m)\mathcal{G}_n.
\label{grozin_massive}
\eea
We note here that after this minimal subtraction with massive correlators
and then taking  the massless limit renders the resulting correlator finite.
Since we are working in a massless limit, one requires an appropriate 
modification~\cite{Broadhurst:1984rr, Broadhurst:1985js} of 
Eq.(\ref{mixed_rep}). The difference between a renormalized quark condensate 
and a bare one can be written as~\cite{Grozin:1994hd},
\bea
\mathcal{Q}_k-\mathcal{Q}_k^{b} = -\frac{1}{\epsilon} \sum_{d_n\le 
d_k}m^{d_k-d_n}\gamma_{kn}\mathcal{G}_n, \label{difference_mixed_rep}
\eea
where $d_n$ and $d_k$ represents the dimensions of $\mathcal{G}_n$ and $\mathcal{Q}_k$  
and $\gamma_{kn}$ are the mixing coefficients of $\mathcal{Q}_k$ with $\mathcal{G}_n$.
Now if one wants to go to $m\rightarrow 0$ limit, $\mathcal{Q}_k^{b}$ vanishes 
because there is no scale involved in it. Also only $d_n=d_k$ term survives producing
\bea
\mathcal{Q}_k = -\frac{1}{\epsilon} \sum_{d_n= 
d_k}\gamma_{kn}\mathcal{G}_n. \label{modified_mixed_rep}
\eea
Using Eq. (\ref{modified_mixed_rep}) in the first line of  Eq.(\ref{grozin_massive}) one 
can write
\bea
\Big[C_{\mu}^{\mu}(p)\Big]_{g}^{\textrm{a}} = 
\Big[C_{\mu}^{\mu}(p)\Big]_{g} + \frac{1}{\epsilon}\sum\limits_{d_n=d_k} 
b_k(p)\gamma_{kn}\mathcal{G}_n, \label{modified_min_sub}
\eea
where, $b_k(p)$ is the coefficient of the quark condensate $\mathcal{Q}_k$ in the 
massless limit, which is of similar dimension as $\mathcal{G}_n$.

So, for minimal subtraction of  the quark condensate contribution 
overestimated in Eq.(\ref{topologyII_massless}), the  in-medium quark condensate 
(appearing in Eq.(\ref{quark_operator_massless})) has to be expressed in terms of the 
in-medium gluon condensates of the same dimension as~\cite{Zschocke:2011aa}
\bea
\big \langle \bar{\psi}\gamma_\mu i D_\nu \psi \big \rangle_T &=& \big \langle : 
\bar{\psi}\gamma_\mu i D_\nu \psi : \big \rangle_T 
+\frac{3}{16\pi^2}m^4 g_{\mu\nu}\left(\ln\frac{\mu^2}{m^2}+1\right)
-\frac{g_{\mu\nu}}{48}\Big \langle \frac{g^2}{4\pi^2}G^2 \Big \rangle_T\nn\\
&&-\frac{1}{18}(g_{\mu\nu}-4u_\mu u_\nu)\left(\ln\frac{\mu^2}{m^2}-
\frac{1}{3}\right)\Big \langle \frac{g^2}{4\pi^2}u\Theta^gu \Big \rangle_T,
\label{hilger}
\eea
where the first term in the right hand side represents the normal ordered condensate.
After contracting Eq.(\ref{hilger}) by $u^\mu u^\nu$ and applying 
Eq.(\ref{quark_condensate_relation1}) we obtain,
\bea
\langle u \Theta^f u \rangle &=& \textrm{Other nonrelevant terms~} + 
\frac{1}{6}\left(\ln\frac{\mu^2}{m^2}-
\frac{1}{3}\right)\Big \langle \frac{g^2}{4\pi^2}u\Theta^gu \Big \rangle . 
\label{qcon_via_gcon}
\eea
Now comparing Eqs.(\ref{qcon_via_gcon}) and (\ref{modified_mixed_rep}) we find 
\bea
\gamma_{kn} = \frac{1}{6},~~b_k(p) 
= \frac{8N_cN_f}{3p^2}\left(1-\frac{4\omega^2}{p^2}\right).\label{coeff_min_sub}
\eea
Therefore, the electromagnetic correlator with gluon condensates for self-energy 
correction (topology-II) in the massless limit can now be written as
\bea
\Big[C_{\mu}^{\mu}(p)\Big]_{g,T}^{\textrm{II,a}}&=\atop{m\rightarrow 0}&\Big[C_{\mu}^{\mu}
(p)\Big ] _ { g , T } ^ {
\textrm{II}}+ \frac{1}{\epsilon}\sum\limits_{d_n=d_k} 
b_k(p)\gamma_{kn}\mathcal{G}_n,\nn\\
 &=& -\frac{g^2N_cN_f}{9\pi^2p^2}\langle u\Theta^gu \rangle_T
\left[\frac{1}{\tilde{\epsilon}}\left(1-\frac{4\omega^2}{p^2}\right)+2-6\frac{\omega^2}{
p^2}\right]+
\frac{1}{\epsilon}\frac{g^2N_cN_f}{9\pi^2p^2}\left(1-4\frac{\omega^2}{p^2}\right)\langle 
u\Theta^gu \rangle_T\nn\\
&=& -\frac{g^2N_cN_f}{9\pi^2p^2}\langle u\Theta^gu \rangle_T 
\left[-\ln\left(\frac{-p^2}{\Lambda^2}\right)\left(1-\frac{4\omega^2}{p^2}\right)+2-6\frac
{\omega^2}{p^2}\right].
\label{minimal_subtracted_topology_II}
\eea
So, the minimal subtraction eventually cancels the divergence from the expression of 
gluonic operators. Now combining Eq.(\ref{topologyI_massless}) and 
Eq.(\ref{minimal_subtracted_topology_II}), the final expression for the gluonic 
contribution in the self-energy power correction is given by,
\bea
\Big[C_{\mu}^{\mu}(p)\Big]_{g,T}\!\!\!\!\!\! \! \! = 
\frac{g^2N_cN_f}{\pi^2p^2}\left[\frac{1}{9}\big\langle u\Theta^gu \big\rangle_T
\left(\ln\left(\frac{-p^2}{\Lambda^2}\right)\!\left(1-\frac{4\omega^2}{p^2}\right)+9\frac{
\omega^2}{p^2}-\frac{11}{4}\right)-\frac{1}{16}\big\langle G^2 \big\rangle_T\right].
\label{gluonic_final}
\eea

\section{Electromagnetic spectral function}
\label{spectral}

The correlation function with power corrections from both quark and gluonic composite 
operators can now be written from Eq.(\ref{quark_operator_massless}) and 
Eq.(\ref{gluonic_final}) as 
\bea
\Big[C_{\mu}^{\mu}(p)\Big]_{T} &=& \Big[C_{\mu}^{\mu}(p)\Big]_{g,T} 
+\Big[C_{\mu}^{\mu}(p)\Big]_{q,T} \nn\\ 
&=&\frac{g^2N_cN_f}{\pi^2p^2}\left[\frac{1}{9}\big\langle u\Theta^gu \big\rangle_T
\left(\ln\left(\frac{-p^2}{\Lambda^2}\right)\!\left(1-\frac{4\omega^2}{p^2}\right)+9\frac{
\omega^2}{p^2}-\frac{11}{4}\right)-\frac{1}{16}\big\langle G^2 \big\rangle_T\right]\nn\\
&+& \frac{8N_cN_f}{3p^2}\big\langle u\Theta^fu 
\big\rangle_T\left(1-4\frac{\omega^2}{p^2}\right)
\left[1+\frac{2g^2}{9\pi^2}\left(1-\ln\left(\frac{-p^2}{\Lambda^2}\right)\right)\right].
\label{correlator_final}
\eea
The contribution to spectral 
function comes from  nonanlytic behavior of $\ln\left(\frac{-p^2}{\Lambda^2}\right)$ 
having a discontinuity of $2\pi$. Following Eq.(\ref{spec_def}) the 
electromagnetic spectral function with leading  non-perturbative power 
corrections in the OPE limit $\pi T<  p< \omega$  can be written as
\bea
\rho^{\textrm{pc}}_f(p)=-\frac{16N_c \alpha_s}{9 
p^2\pi}\left(1-4\frac{\omega^2}{p^2}\right)\left[\frac{8}{3}\Big\langle \Theta^f_{00} 
\Big\rangle_T-\frac{1}{2}\Big\langle \Theta^g_{00} \Big\rangle_T\right],
\label{spec_func_pc}
\eea
where the power corrections ($p^{D/2}$) from 
the QCD vacuum, resides in the denominator of the  Wilson coefficient as we have 
considered the $D=4$ dimensional composite operators. Now, $\Theta_g^{00}$ and 
$\Theta_f^{00}$ are respectively the gluonic and fermionic part 
of the energy density $\mathcal{E}$, and in the Stefan-Boltzmann limit given by
\bea
\Big\langle \Theta^g_{00} \Big\rangle^{\textrm{SB}}_T &=& \frac{\pi^2T^4}{15}d_A,\nn\\
\Big\langle \Theta^f_{00} \Big\rangle^{\textrm{SB}}_T &=& 
\frac{7\pi^2T^4}{60}d_F,\label{opera_sb}
\eea
where, $d_A=N_c^2-1$ and $d_F=N_cN_f$. So the leading correction to the 
electromagnetic spectral function is ${\cal O}(g^2T^4)$ and  is in conformity 
with those obtained in Ref.\cite{CaronHuot:2009ns} using renormalization group equations 
(RGE). Now the perturbative leading order (PLO)  result, or the free spectral 
function, is given by
\bea
\rho^{\textrm{PLO}}(p)=\frac{N_c N_f Tp^2}{4 \pi^2 
\vert\vec{p}\vert}\ln\left[\frac{\cosh\left(\frac{\omega + 
\vert\vec{p}\vert}{4T}\right)}{\cosh\left(\frac{\omega - 
\vert\vec{p}\vert}{4T}\right)}\right].
\label{spec_func_lop}
\eea
Now to have a quantitative estimates of the physical quantities considered, one
needs the in-medium values of those condensates appearing in Eq.(\ref{spec_func_pc}) in 
region of interest ($\pi T< p<\omega$).  Unfortunately, the present knowledge of those 
in-medium condensates  are not available in the existing literature.
The evaluation 
of the composite quark and gluon operators (condensates) in Eq.(\ref{spec_func_pc}),  
$\langle \Theta_g^{00}\rangle_T$ and $\langle \Theta_f^{00}\rangle_T$, respectively, at 
finite $T$  should proceed via nonperturbative methods of QCD. 
We expect that LQCD calculations would be able to provide some preliminary estimate of 
them in near future. Since presently we do not have any information of these in-medium 
condensates, we just use their Stefan-Boltzman limits, as given in Eq.(\ref{opera_sb}), 
to have some limiting or qualitative information even though it is not appropriate at the 
region of 
interest. For quantitative estimates one should  wait until actual estimates of these 
condensates, $\langle \Theta_g^{00}\rangle_T$ and $\langle \Theta_f^{00}\rangle_T$, are 
made avilable in the literature.

Also we use the one-loop running coupling
\bea
\alpha_s(\Lambda)&=&\frac{12\pi}{33-2N_f}\, \ln\frac{\bar\Lambda_{\rm 
MS}^2}{\Lambda^2},
\eea
with  $\bar\Lambda_{\rm MS}=176{\rm MeV}$ \cite{alphas1} and the 
renormalisation scale is chosen at its central value, $\Lambda=2\pi T$.

\begin{center}
\begin{figure}[h]
 \begin{center}
 
\includegraphics[scale=1.0]{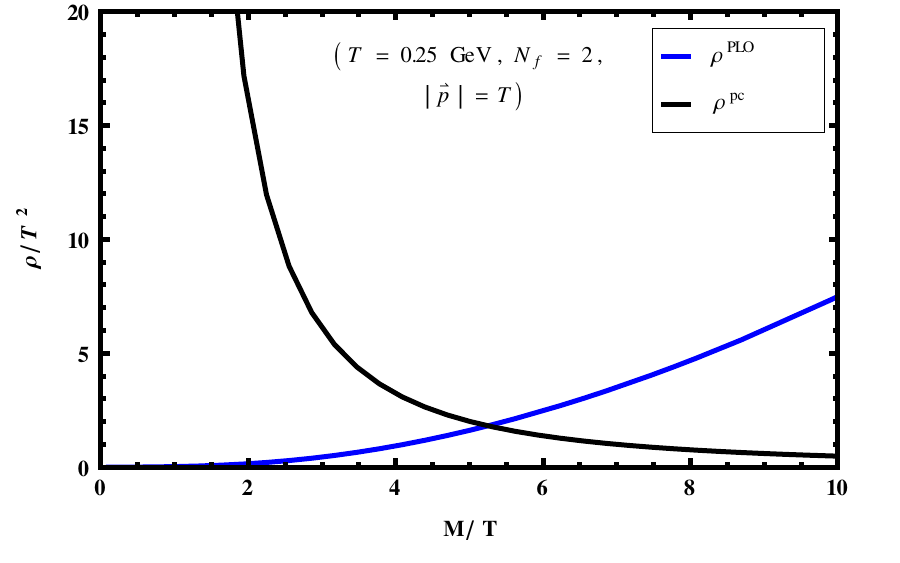}%\includegraphics[scale=0.9]{
%spectral_ratio.pdf}
 \caption{Comparison of the  electromagnetic spectral function betwen  
the perturbative leading order (PLO)  and power corrections from $D=4$.}
  \label{spec_plot}
 \end{center}
\end{figure}
\end{center}

We now demonstrate the importance of the power corrections  in 
the thermal spectral function. Figure \ref{spec_plot} displays a comparison between 
the perturbative leading order contribution in Eq.(\ref{spec_func_lop}) and the power 
corrections contribution in Eq.(\ref{spec_func_pc}). As seen that the nature of 
the two spectral functions are drastically different to each other as a function
of $M/T$, the scaled invariant mass with respect to temperature. While the perturbative 
leading order result increases with the increase of $M/T$, the leading order 
power correction starts with a very high value but falls off very rapidly. We, here, 
emphasize that the low invariant  mass region is excluded in the OPE  limit, 
$\pi T < p <\omega$. On the other hand the vanishing contribution of the power 
corrections 
at large invariant mass ($M\approx 10T$) is expected because of the appearance of 
$p^{-2}$ due to dimensional argument as discussed after  Eq.(\ref{spec_func_pc}). So, at 
large invariant mass the perturbative calculation becomes more effective as can be seen 
in Fig.\ref{spec_plot}. 

\begin{center}
\begin{figure}[h]
 \begin{center}
 
\includegraphics[scale=0.8]{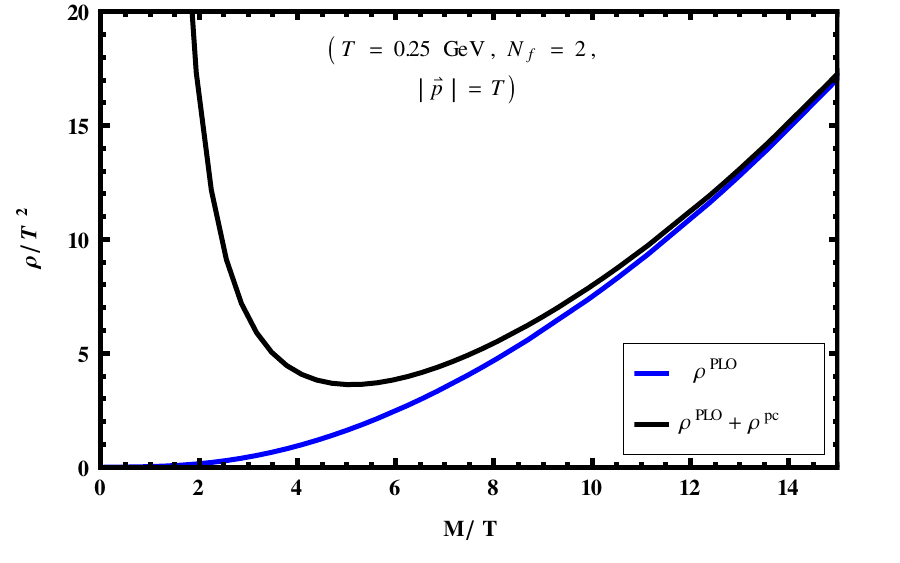}
\includegraphics[scale=0.8]{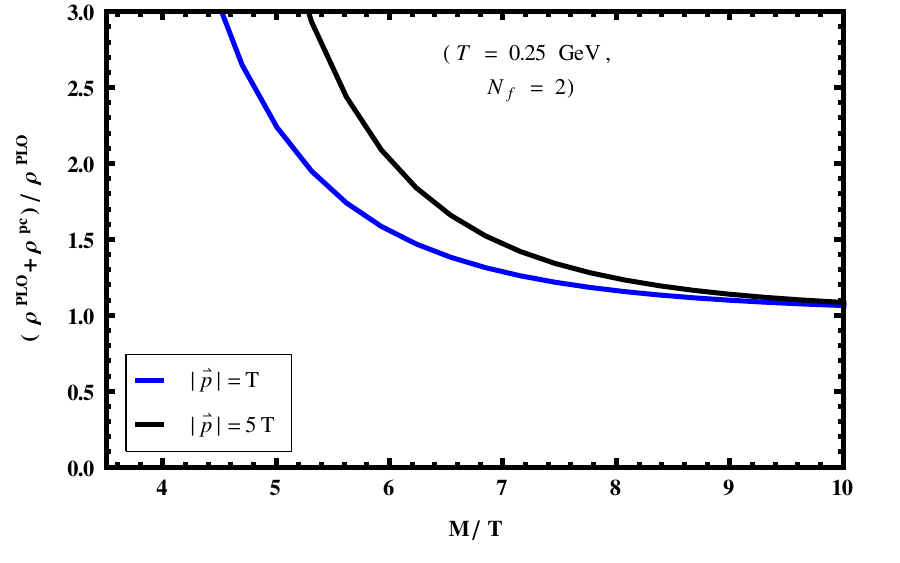}
 \caption{ Comparison between  $\rho^{\textrm{PLO}}$ in Eq.(\ref{spec_func_lop}) and   
$\rho^{\textrm{PLO}} + \rho^{\textrm{pc}}$ in the \textit{left panel}  and  their ratio 
in the \textit{right panel}. }
  \label{spec_plot_2}
 \end{center}
\end{figure}
\end{center}

In figure \ref{spec_plot_2}, a comparison (left panel) between the $\rho^{\textrm{PLO}}$ 
in Eq.(\ref{spec_func_lop}) and   
$\rho^{\textrm{PLO}} + \rho^{\textrm{pc}}$ 
[Eq.(\ref{spec_func_lop})+Eq.(\ref{spec_func_pc})] and their ratio are displayed, 
respectively. From the 
left panel one finds that in the intermediate mass regime,  $M\approx 4T$ to  $10T$, 
\textrm{i.e.}, (1 to 2.5) GeV, there is a clear indication of enhancement in the  
electromagnetic spectral function due to the leading order power corrections in $D=4$ 
dimension.  This is also reflected in the ratio plot in the right panel. Both plot 
assures that the power corrections becomes important in the intermediate mass range of 
the electromagnetic spectral function. 

\begin{center}
\begin{figure}[h]
 \begin{center}
\includegraphics[scale=0.9]{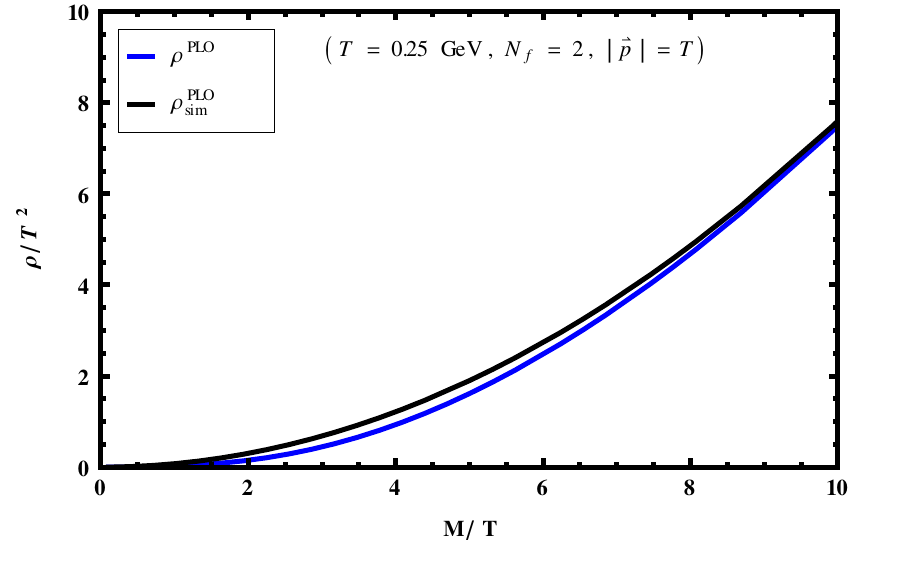}
 \caption{ Comparison between  $\rho^{\textrm{PLO}}$ in Eq.(\ref{spec_func_lop}) and  
simplified one  $\rho^{\textrm{PLO}}_{\textrm{sim}}$ in Eq.(\ref{plo_ope}).}
\label{plo_ope_limit}
 \end{center}
\end{figure}
\end{center}

For convenience the PLO spectral function in Eq.(\ref{spec_func_lop}) can be simplified 
in the OPE limit as
\bea
\rho^{\textrm{PLO}}_{\textrm{sim}}(p)&=&\frac{N_cN_fp^2}{4 \pi^2}. \label{plo_ope}
\eea
and is also justified through Fig.~\ref{plo_ope_limit}.

The total spectral function with  the power 
correction in the OPE limit can now be written as
\bea
\rho(p)\vert^{OPE}&=& \rho^{\textrm{PLO}}_{\textrm{sim}}(p)+\rho^{\textrm{pc}}(p) \nn \\
&=& \frac{N_cN_fp^2}{4 \pi^2}-\frac{16N_cN_f \alpha_s}{9\pi 
p^2}\left(1-4\frac{\omega^2}{p^2}\right)\left[\frac{8}{3}\Big 
\langle \Theta_f^{00}\Big \rangle_T -\frac{1}{2} \Big \langle\Theta_g^ { 
00}\Big \rangle_T\right
] .
\label{spec_func_ope}
\eea
Now we note that the virtual photon will decay into two leptons and the features 
observed in the electromagnetic spectral function will also be reflected in the dilepton 
production rate, which will be discussed in the next 
section. 

\section{Dilepton Rate}\label{dilepton_rate}

The modified PLO  differential dilepton production rate in presence of leading power 
correction in OPE with $D=4$  is now obtained in a compact form by combining 
Eqs.(\ref{dilepton_rate_final}) and (\ref{spec_func_ope}) for $N_f=2$ as
\bea
\label{dilepton_rate_final}
{\frac{dR}{d^4xd^4p}}\Bigg\vert^{\textrm{OPE}} &=&  \frac{5\alpha^2_{\textrm 
em}}{27 \pi^2 M^2}
n_B\left(\omega\right) \nn \\ 
&&\times \left[ \frac{N_cp^2}{4 \pi^2} -\frac{16N_c \alpha_s}{9\pi 
p^2}\left(1-4\frac{\omega^2}{p^2}\right)\left(\frac{8}{3}\Big 
\langle \Theta_f^{00}\Big \rangle -\frac{1}{2} \Big \langle\Theta_g^ { 
00}\Big \rangle\right ) \right ],
\eea
where we have used $\sum_f q_f^2 =5/9$, for massless $u$ and $d$ quarks.
The leading power corrections within  OPE in $D=4$ dimension is of ${\cal 
O}(\alpha^2_{em} \alpha_s)$ to the PLO of ${\cal O}(\alpha^2_{em})$.

\begin{center}
\begin{figure}[h]
 \begin{center}
 \includegraphics[scale=1.]{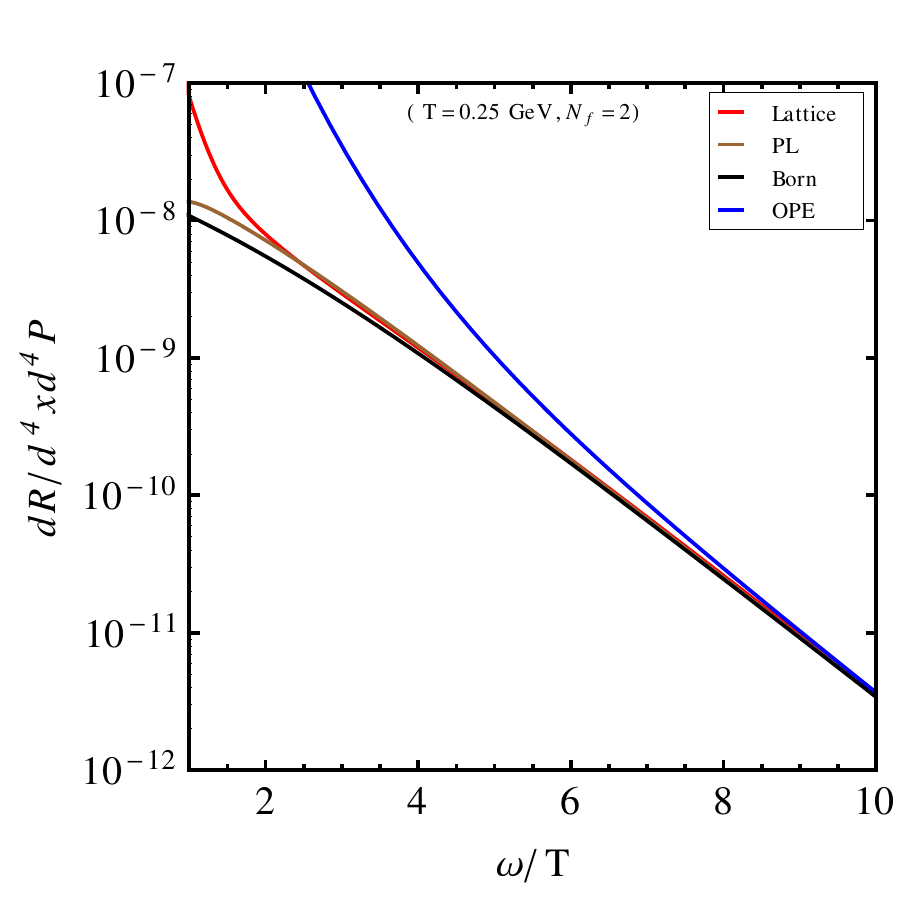} 
 \caption{Comparison between different dilepton rates as a function of  $\omega/T$ with 
$T=250$ MeV, obtained from Lattice simulations~\cite{Ding:2010ga,Ding:2016hua}, PL based  
model 
calculations~\cite{Islam:2014sea,Gale:2014dfa}, Born rate and the nonperturbative power 
corrections}
  \label{dilepton_final}
 \end{center}
\end{figure}
\end{center}

In  Fig. \ref{dilepton_final}, a comparison is displayed among various thermal dilepton 
rates as a function of $\omega/T$ with $T=250$ MeV and zero external three momentum. 
The various dilepton rates considered here are  Born (PLO)~\cite{Greiner:2010zg, 
Cleymans:1992gb}, PLO plus power 
corrections within OPE in Eq.(\ref{dilepton_rate_final}), 
LQCD~\cite{Ding:2010ga,Ding:2016hua} and 
Polyakov Loop (PL) based models in an effective QCD 
approach~\cite{Islam:2014sea,Gale:2014dfa}. The dilepton rate from  PL based models
and LQCD  for 
$\omega/T>4$ becomes simply perturbative in nature whereas  it is so for 
$\omega/T \ge 10$ 
in case of the PLO with power corrections in OPE. For $\omega/T>4$, in PL based  models 
the confinement effect due to Polyakov loop becomes very weak whereas in LQCD the 
spectral 
function is replaced by the PLO one. On the other hand the enhancement of the dilepton 
rate  at low energy ($\omega/T < 4$) for both PL based models and LQCD is due to the
presence of some nonperturbative effects ({\textrm{e.g.,}} residual confinement effect 
etc) whereas in that region the power corrections within OPE is not 
applicable\footnote{In principle one can approximate the dilepton rate in the low mass, 
$\omega/T \le 4$, region by the results from perturbative next-to-LO (PNLO) 
~\cite{Laine:2013vma, Ghisoiu:2014mha, Ghiglieri:2014kma} and 1-loop hard thermal loop 
(HTL) resummation~\cite{Braaten:1990wp}, which agree to each other in order of 
magnitudes. We also note that in the low mass regime (soft-scale)
the perturbative calculations break down as the loop expansion has its generic 
convergence problem in the limit of small coupling ($g \le 1$). On the other hand  PLO, 
PNLO,  HTL resummation and OPE agree in the hard scale, \textit{i.e.}, in the very large 
mass $\omega/T\ge 10$.}. However, in the intermediate domain ($4 <\omega/T < 10 $) the 
dilepton rate is enhanced compared to PL based models and LQCD due to the
presence of the nonperturbative composite quark and gluon operators that incorporates 
power corrections within OPE in $D=4$.  We note that 
the power corrections in OPE considered here may play an important role for intermediate
mass dilepton spectra from high energy heavy-ion collisions in RHIC and LHC. 

\section{Conclusion}
\label{conclusion}

QCD vacuum has a nontrivial structure due to the fluctuations of the quark and gluonic 
fields which generate some local composite operators of quark and gluon fields, 
phenomenologically known as condensates. In perturbative approach by definition such 
condensates do not appear in the observables. However, the nonperturbative dynamics of 
QCD is evident through the power corrections in physical observables by considering the 
nonvanishing vacuum expectation values of such local quark and gluonic composite 
operators.
In this paper we, first, have made an attempt to compute the nonperturbative 
electromagnetic spectral function in QCD plasma by taking into account the power 
corrections and the nonperturbative condensates within the framework of the OPE 
in $D=4$ dimension. The power corrections appears in the in-medium 
electromagnetic spectral function through the nonanalytic behavior 
of the current-current correlation function in powers of $p^{-D/2}$ or logarithms in the 
Wilson coefficients within OPE in $D=4$ dimension. The $c$-numbered Wilson coefficients 
are computed through Feynman diagrams by incorporating the various condensates. 
In the massless limit of quarks, the self-energy diagram involving local gluonic operator 
(topology-II in Fig.\ref{topology_2}) encounters mass singularity. By exploiting the 
minimal subtraction through operator mixing this mass singularity cancels out, which 
renders the Wilson coefficients free from any infrared singularity and hence finite.
This result is in conformity with the RGE analysis.

The lepton pairs are produced through the electromagnetic interaction 
in every stage of the hot and dense medium created in high energy 
heavy-ion collisions. They are considered to be an important probe of QGP formation 
because  they leave, immediately after their production,  the hot and dense medium 
almost without any interaction.
As a spectral property of the electromagnetic spectral function, we then
evaluated  the differential dilepton production rate from QCD plasma in the 
intermediate mass range to analyze the effects of power corrections and nonperturbative 
condensates. The power correction contribution is found to be  ${\cal 
O}(\alpha^2_{\textrm{em}}\alpha_s)$ to the PLO, ${\cal 
O}(\alpha^2_{\textrm{em}})$. Further, we note that the intermediate mass range is 
considered because the low mass regime ($ M \le 4T \sim 1$ GeV; $T=0.25$ GeV) is 
prohibited by OPE whereas high mass regime ($ M \ge 10T \sim 2.5$ GeV) is well described 
by the perturbative approach. The intermediate mass range ($4T\le M \le 10T$) dilepton 
in presence of power corrections is found to be enhanced compared to other nonpertubative 
approaches, \textit{i.e.}, LQCD and 
effective QCD models. However,  we note that the 
power corrections in differential dilepton rate through OPE considered here could be 
important to describe the intermediate mass dilepton spectra from heavy-ion collisions. 

Finally,  we would like to note that there is no estimate available in the 
present literature for the composite quark and gluon operators (condensates),  
$\langle \Theta_g^{00}\rangle_T$ and $\langle \Theta_f^{00}\rangle_T$, respectively, at 
finite $T$. Since the present knowledge of these in-medium operators are very meagre in 
the literature, we have exploited  the Stefan-Boltzmann limits for these composite 
operators to have some limiting information of the nonperturbative effects in the 
electromagnetic spectral function and its spectral properties. We expect that in near 
future the computation of such phenomenological quantities should be possible via 
nonperturbative methods of QCD in lattice and some definite estimation of 
the power corrections within OPE can only be made for spectral function and its spectral 
properties.

\section{Acknowledgements}

The authors would like to acknowledge useful discussions with S. Leupold,  S. Mallick, C. 
A. Islam and 
N. Haque. AB would specially like to thank  P. Chakrabarty  for very enlighting 
discussion. This work is supported by Department 
of Atomic Energy (DAE), India under the project $``$Theoretical Physics Across
The Energy Scale (TPAES)'' in Theory Division of Saha Institute of Nuclear Physics.

\appendix 

\section{Appendix}
\subsection{Massless Feynman Integrals}\label{massless_feynman}

While computing the electromagnetic polarization tensor with gluon condensates, the 
following Feynman integrals for massless quarks have been used: 
\bea
\mathcal{I}_{mn}  &=& \int \frac{d^dk}{(2\pi)^d} \frac{1}{(k^2)^m((k-p)^2)^n},\nn\\
\mathcal{I}_{mn}^\mu &=& \int \frac{d^dk}{(2\pi)^d} \frac{k^\mu}{(k^2)^m((k-p)^2)^n},\nn\\
\mathcal{I}_{mn}^{\mu\nu}  &=& \int \frac{d^dk}{(2\pi)^d} \frac{k^\mu 
k^\nu}{(k^2)^m((k-p)^2)^n}.\nn
\eea
The primary integrals can be represented as follows,
\bea
\mathcal{I}_{mn} &=& \frac{i}{(16 
\pi^2)^{\frac{d}{4}}}(-1)^{-m-n}(-p^2)^{-m-n+\frac{d}{2}}
\frac{\Gamma[m+n-\frac{d}{2}]}{\Gamma[m]\Gamma[n]}B\left(\frac{d}{2}-n,
\frac{d}{2}-m\right),\\
\mathcal{I}_{mn}^\mu &=& \frac{i}{(16 
\pi^2)^{\frac{d}{4}}}(-1)^{-m-n}(-p^2)^{-m-n+\frac{d}{2}}\nn\\
&&~~~~~~~~~~~~~ p^\mu \Biggl\{\frac{\Gamma[m+n-\frac{d}{2}]\Gamma[1+\frac{d}{2}-m]\Gamma[
\frac{d}
{2}-n]}
{\Gamma[m]\Gamma[n]\Gamma[1+d-m-n]}\Biggr\},\\
\mathcal{I}_{mn}^{\mu\nu} &=& \frac{i}{(16 
\pi^2)^{\frac{d}{4}}}(-1)^{-m-n}(-p^2)^{-m-n+\frac{d}{2}}\nn\\
&&~~~~~~~~~~~~~\Biggl\{p^2g^{\mu\nu}\frac{\Gamma[m+n+2-\frac{d}{2}]\Gamma[1+\frac{d}{2}-m]
\Gamma[1+\frac{d}{2}-n]}
{2\Gamma[m]\Gamma[n]\Gamma[2+d-m-n]}\nn\\
&&~~~~~~~~~~~~~~+p^\mu 
p^\nu\frac{\Gamma[m+n-\frac{d}{2}]\Gamma[2+\frac{d}{2}-m]\Gamma[\frac{d}{2}-n]}
{2\Gamma[m]\Gamma[n]\Gamma[2+d-m-n]}\Biggr\}.
\eea
Now putting $d = 4 - 2 \epsilon$, we obtain required results of $\mathcal{I}_{mn}, 
\mathcal{I}_{mn}^\mu$ and $\mathcal{I}_{mn}^{\mu\nu}$ 
for some given values of $m$ and $n$ needed for our purpose: 
\bea
\mathcal{I}_{12} &=& 
\mu^{-\epsilon}\frac{i}{16\pi^2}\frac{1}{p^2}\left(-\frac{1}{\tilde{\epsilon}}\right),
\nn\\
\mathcal{I}_{22} &=& 
\mu^{-\epsilon}\frac{i}{16\pi^2}\frac{1}{p^4}2\left(-\frac{2}{\tilde{\epsilon}}-2\right),
\nn\\
\mathcal{I}_{22}^0 &=& 
\mu^{-\epsilon}\frac{i}{16\pi^2}\frac{p^0}{p^4}\left(-\frac{1}{\tilde{\epsilon}}-1\right),
\nn\\
\mathcal{I}_{31}^0 &=& 
\mu^{-\epsilon}\frac{i}{16\pi^2}\frac{p^0}{p^4}\left(-\frac{1}{2\tilde{\epsilon}}-\frac{1}
{2}\right),\nn\\
\mathcal{I}_{22}^{00} &=& 
\mu^{-\epsilon}\frac{i}{16\pi^2}\frac{1}{p^4}\left[\frac{p^2}{2}+\left(-\frac{1}{\tilde{
\epsilon}}-2\right)(p^0)^2\right],\nn\\
\mathcal{I}_{31}^{00} &=& \mu^{-\epsilon}\frac{i}{16\pi^2}\frac{1}{(p^2)^2}\left[p^2
\left(-\frac{1}{4\tilde{\epsilon}}-\frac{1}{4}\right)+\frac{(p^0)^2}{2}\right],\nn\\
\mathcal{I}_{41}^{00} &=& \mu^{-\epsilon}\frac{i}{16\pi^2}\frac{1}{(p^2)^3}\left[p^2
\left(\frac{1}{12\tilde{\epsilon}}-\frac{1}{12}\right)+(p^0)^2\left(-\frac{1}{3\tilde{
\epsilon}}-\frac{1}{2}\right)\right],\nn
\eea
where
\bea
\mu &=& e^{\frac{\gamma_E}{2}}\frac{\Lambda^2}{4\pi},\nn\\
\frac{1}{\tilde{\epsilon}} &=& 
\frac{1}{\epsilon}-\log\left(\frac{-p^2}{\Lambda^2}\right),\nn
\eea
with $\mu$ as the renormalization scale, $\Lambda$ as $\overline{\rm MS}$ renormalization 
scale and $\gamma_E$ as Euler-Mascheroni constant.

\end{document}